\documentclass[prc,twoside,floatfix,superscriptaddress,showkeys,showpacs,twocolumn]{revtex4}

\usepackage{graphicx, latexsym, amssymb, amsmath, color, multirow, mathrsfs, CJK, ifpdf}
\usepackage[section]{placeins}
\usepackage{amsmath}
\usepackage{amsfonts}
\usepackage{amssymb}
\usepackage{bm}
\usepackage{graphicx}
\usepackage{epstopdf}
\usepackage{color} 
\usepackage{txfonts}
\usepackage{ulem}

\makeatletter

\newcommand{\Rmnum}[1]{\expandafter\@slowromancap\romannumeral#1@}
\makeatother
\usepackage{dcolumn}   
\usepackage{physics}
\usepackage[breaklinks=false]{hyperref}
\begin{document}

\preprint{APS/123-QED}

\title{Nuclear level density studied in odd-mass nuclei in the framework of projected shell model}

\author{Jiaqi Wang}
\affiliation{School of Physics and Astronomy, Shanghai Jiao Tong University, Shanghai 200240, China}
\author{Saumi Dutta}
\author{Cui-Juan Lv}
\affiliation{School of Physics and Astronomy, Shanghai Jiao Tong University, Shanghai 200240, China}
\author{Long-Jun Wang}
\affiliation{School of Physical Science and
Technology, Southwest University, Chongqing 400715, China}
\author{Yang Sun}\email{sunyang@sjtu.edu.cn  (correspondence)}
\affiliation{School of Physics and Astronomy, Shanghai Jiao Tong University, Shanghai
200240, China}

\date{\today}
\begin{abstract}
  
In a recent article (Phys. Rev. C {\bf 108}, 034309 (2023)), we proposed a projected shell model method for the calculation of nuclear level density (NLD) in deformed even-even nuclei. The current article presents the subsequent study of NLDs in odd-mass nuclei as well as a comparative analysis between our calculated NLDs in adjacent even-even and odd-$A$ systems. Since one nucleon in the odd-mass system remains blocked from participating in the pair formation, resulting in a weakened pairing (assessed by smaller BCS pairing gap $\Delta$), pronounced differences between the NLDs in an odd-mass (both even-odd and odd-even) nucleus and its immediate even-even neighbor have been found. In general, the structure-dominant variations, which were found to be prominent in the even-even NLD at low energies, are greatly suppressed in the odd-mass systems. Specifically, from excitation energy as low as 2 MeV, the calculated densities of odd-parity and even-parity levels in odd-mass nuclei show an equal division signaling faster attainment of the statistical behavior. Nuclear level-spin distributions of both parities have been seen to adopt a regular Gaussian shape earlier than that found in the even-even system. Moreover, the pleasant property of our shell-model results, that each of our calculated levels is an eigenstate of angular momentum, allows us to extract the values of the energy-dependent dispersion $\sigma$ of Ericson's spin-distribution formula and plot  $\rho (E_x, I, \pi)$, the energy, spin, and parity dependent level density. 
\end{abstract}


\maketitle

\section{Introduction}\label{intro}

Nuclear reaction and nuclear structure are two different, yet closely related branches of nuclear science. For studies including astrophysics and several practical applications, such as reactor technology, nuclear medicine, and nuclear waste management, radiative neutron capture is one of the important types of reactions. The common description of this reaction process \cite{Larsen2019} relies on the Hauser-Feshbach theory \cite{Hauser-Feshbach} which requires the nuclear optical potential, nuclear level density (NLD), and $\gamma$-ray strength function as model inputs. The latter two, the key ingredients in the Hauser-Feshbach formula for determining reaction cross-sections, are purely nuclear structure quantities. 

It is therefore critically important to get information on high-quality nuclear level densities and $\gamma$-ray strengths. Among different existing experimental techniques, the Oslo method \cite{Oslo} is a powerful tool to simultaneously extract nuclear level density and $\gamma$-ray strength function for the energy spectrum below the particle threshold. The Oslo method, together with its two parallel extensions, namely, the $\beta$-Oslo method \cite{Beta-Oslo} and the inverse Oslo method \cite{Inverse-Oslo}, has produced a wealth of nuclear level densities and strength functions for stable as well as unstable nuclei over a wide range of excitation energy \cite{Goriely_EPJA}. However, the final step of the Oslo method, at which the common slope of NLD and $\gamma$-ray strength function is determined, requires absolute normalization with the experimentally known data, which introduces some model dependency. Recently, the shape method \cite{Shape method, Shape_method_1} has been introduced, which, as mostly a model-independent approach, offers an advantage over the conventional Oslo method in the sense that the slope of the $\gamma$-ray strength function and the NLD can be determined without the need of experimental auxiliary data for normalization. For the cases which cannot be probed by experiments, information must be obtained from reliable theoretical calculations.

While being implemented as structural inputs in the Hauser-Feshbach theory \cite{Hauser-Feshbach}, NLDs and $\gamma$-ray strengths are required to be functions of nuclear state properties. For example, as far as NLD is concerned, it is needed as functions of excitation energy, spin, and parity, i.e., $\rho (E_x, I, \pi)$. However, the above-mentioned experimental methods do not immediately provide densities (within the excitation energy bins of fixed widths) of the levels sorted according to definite spin and parity values, but rather as a quantity, $\rho(E_x)$, in which levels of all spins and parities are mixed altogether. Therefore, while using such NLDs in various calculations, one has to assume that spin and parity are described by uncorrelated functions, as follows:
\begin{equation}\label{EQ: uncorrelated_NLD}
\rho (E_x, I, \pi) = \rho(E_x) ~g(E_x, I) ~{\cal F}(E_x, \pi),
\end{equation}
where, $g$ is the spin-distribution and $\cal F$ is the parity-distribution. Finding a proper expression of spin-distribution in NLDs has been a long-standing problem. Without experimentally determined spin, the commonly adopted model for $g$ is that of Ericson \cite{Ericson1959} based on the statistical random coupling of angular momentum. However, in addition to its shortcomings at low excitation energy due to the lack of detailed nuclear structure, the validity of Ericson's expression is in question for very large values of $I$ \cite{Larsen2019}. 

Therefore, it is clear that theoretical modeling is indispensable not only to find reasonable descriptions of NLDs but also to get information on the appropriate distribution of nuclear level-spins. In the pioneering work of Bethe \cite{Bethe1937}, the level density was described within the concept of the Fermi gas model using a partition function approach for the grand canonical ensemble, in which nuclear energy levels were determined by summing up the energies of the non-interacting particles. This much-simplified concept, which completely neglected the shell and pairing effects, was later subjected to several empirical corrections to account for the missing effects giving rise to phenomenological models as extensions to the original Fermi gas model. Among them, the back-shifted Fermi gas model \cite{Gilbert-Cameron} is the most well-known and widely used extension. Since these models do not contain any structural information, they cannot contribute to the solution of spin-distribution. Modern microscopic models \cite{combinatorial_hfb_0,combinatorial_hfb_1} are based on the concept of cumulative counting of possible ways to distribute the nucleons among the single-particle (SP) levels associated with the Hartree-Fock-Bogolyubov (HFB) mean-field. Consequently, the important shell effects, pair correlations, and deformation can be realized in a self-consistent way \cite{Ring-Schuck}. In this sense, the (deformed) single particles in these models are more physical. However, angular momentum is generally violated in the variation calculation for deformed nuclei, and therefore, the obtained states are not angular momentum states. Hence, neither the phenomenological models nor the microscopic models based on the mean-field theory can provide information on the distribution of nuclear level spins.

Among the other existing theories for nuclear level density, the shell model Monte Carlo (SMMC) method \cite{monte-carlo-shell-model,smmc_2,smmc_3} enables the exact calculation (up to statistical errors) of NLDs. The SMMC allows one to use many-particle model spaces that are many orders of magnitude larger than those that the conventional shell model methods can handle. In the original SMMC approach, thermal averages are taken over all possible states of a given nucleus, and thus the computed level densities are summed over all possible spin states. Later, in Ref. \cite{Alhassid2007}, Alhassid {\it et al.} introduced an approximate angular momentum projection method into SMMC so that the thermal observables can be obtained at given spins.

It has become clear that a breakthrough in nuclear many-body theory calculations is required to obtain high-definition nuclear level density values as functions of explicit level-spins and level-parities. We have recently introduced a novel shell model method to calculate nuclear level density \cite{jiaqi2023}. The method is based on the projected shell model (PSM) \cite{PSM_Hara_Sun, PSM_Sun}, which adopts a large SP space and physically selected many-body configurations in terms of broken pairs. Within the truncated configuration space guided by physics, it is possible to solve the many-body eigenvalue problem and obtain all eigenvalues and eigenstates for arbitrarily heavy and deformed nuclei. NLDs with definite spin/parity can then be naturally constructed from the solution. By applying this method, in Ref.~\cite{jiaqi2023}, we presented the NLD distribution in even-even $^{164}$Dy nucleus along with the parity- and spin-dependence in the calculated levels. Not only did we find a quantitative agreement with experimentally known low-energy discrete levels, but also predicted a step-like structure in $^{164}$Dy NLD curve that had not been discussed before. We attributed this new structure to the contribution of special four-quasiparticle (4-qp) states made of a broken neutron pair plus a broken proton pair, a phenomenon unique to the fermionic nuclear system having two components of isospin.  

In odd-mass (denoted as odd-$A$ hereafter) nuclei, one nucleon remains singled out from pair formation after the rest all get paired together. Therefore, to compare with even-even nuclei, odd-$A$ systems have unpaired nucleons already in the ground state. With this distinct feature, odd-$A$ nuclei can pose various new phenomena related to NLD, which are not found in our preceding study for the even-even counterparts \cite{jiaqi2023}. In the shell model approach, the calculated NLD strongly depends on the distribution of SP orbitals near the Fermi level. Usually, a few (and up to about ten) deformed 1-qp states exist around the Fermi levels in an odd-$A$ nucleus. Therefore, depending on the deformed SP structure, the density of levels in an odd-$A$ nucleus should be, on average, a few to ten times higher than those in their neighboring even-even adjoins at the same excitation energy. This pronounced difference between odd-$A$ and even-even systems can lead to interesting consequences which will be addressed in the present article using our PSM results. 

Compared to even-even nuclei, fewer NLD studies exist for odd-$A$ ones. Quite a long time ago, in 1989, experimentalists at Grenoble, Riga, and Brookhaven were able to obtain a level scheme for $^{163}$Dy up to 2 MeV including 277 $\gamma$ transitions and 14 rotational band assignments \cite{data-Dy163}. The modern Oslo facility could provide NLDs in some selected odd-$A$ nuclides up to the vicinity of neutron separation energy. On the theoretical side, very few studies and discussions exist for the odd-$A$ nuclei. The aforementioned phenomenological models or the microscopic model based on the mean-field theory do not generally distinguish the odd-$A$ from the even-$A$ systems. Application to odd-$A$ NLDs by the SMMC was hampered by the sign problem originating from the projection on an odd number of particles. In an attempt to circumvent this problem, in Ref. \cite{smmc_odd_mass}, Mukherjee and Alhassid discussed a way to distinguish the SMMC calculation for the odd-particle-count system from the even-particle-count one. However, in real applications \cite{smmc_rare_earth_odd-even}, certain empirical considerations need to be introduced within such SMMC approach for determining the odd-$A$ NLDs. Thus, the question of whether the existence of an unpaired nucleon in odd-$A$ nuclei can control and decide energy-, parity-, and spin-dependence of nuclear levels has not been well explored.

The present paper discusses the NLD calculation for the odd-$A$ nuclei, taking rare-earth $^{163}$Dy and $^{163}$Tb nuclei as candidates, which are, respectively, isotope and isotone of $^{164}$Dy nucleus that we used in our previous study as the representative of the even-even system \cite{jiaqi2023}. The choice of such a group of neighboring nuclides would help to investigate the differences between the density of nuclear levels in adjacent even-even and odd-$A$ systems as well as the systematics of level-spin and level-parity distributions in them.

The paper is arranged as follows. We have introduced the projected shell model method for odd-$A$ nuclei in Section \ref{PSM_theory} where we have noted down the qp configuration spaces distinctly for even-odd and odd-even cases, that are used within PSM. We have presented the results of our PSM calculations with detailed discussions in Section \ref{PSM_results}. Before presenting our results of NLD, first, in Section \ref{discrete_level}, we have validated the applicability of our model for both even-odd and odd-even systems through a quantitative comparison of the obtained discrete levels for $^{163}$Dy and $^{163}$Tb with the available data from spectroscopic measurements. Structure-dominant NLDs in $^{163}$Dy and $^{163}$Tb as the function of the excitation energies are presented in Section \ref{NLDs}, where the intriguing process of how the multi-qp configurations of leading orders contribute to the exponential rise of the NLD curves is delineated. In Section \ref{parity}, we have discussed the parity-distribution in NLDs of $^{163}$Dy and $^{163}$Tb and compared the results with their immediate even-even neighbor $^{164}$Dy. Spin distribution in levels of odd-$A$ nuclei is analyzed in Section \ref{spin-distribution}. We have demonstrated how we extract the value of the dispersion, commonly referred to as $\sigma$, of Ericson's spin-distribution formula \cite{Ericson1959} for levels of separate parities. From our calculations, we present NLDs as the function of definite angular momentum from which differences in the spin-dependence of even-even and odd-$A$ NLDs can be clearly viewed. Finally, summary and future prospective are given in Section \ref{Summary}.

\section{Projected shell model for odd-$A$ nuclei}\label{PSM_theory}

The idea of angular momentum projection for odd-$A$ systems was first implemented by Hara and Sun in the shell model configuration mixing calculations \cite{Hara1992}. Later, it was developed as a part of the PSM-family for odd-$A$ nuclear systems \cite{Sun1997}, and considered to be an efficient means to realize shell model calculations for such systems with one unpaired nucleon \cite{PSM_Sun}. Since then, along with many examples of even-even and odd-odd nuclei, this shell model method has been applied to understand various structure observations found in odd-$A$ systems. In the early days, rotational properties of odd-$A$ nuclei found through nuclear high-spin spectroscopy were extensively studied \cite{Sun1994a, Sun1994b, Ta177-Exp}. For example, Ref. \cite{Sun1994a} explained the puzzling observation of the much-delayed band-crossing frequency which, at one time, was thought as an anomaly. In recent years, the application of the PSM method to odd-$A$ nuclei has been successfully extended, notably to various structure problems having nuclear astrophysics relevance, such as the calculation of stellar electron capture rates and Gamow-Teller transition rates \cite{EC_rates, urca_cooling, Wang_2018_GT_rates}. In Ref. \cite{urca_cooling}, Wang {\it et al.} emphasized the effect of nuclear excitations in the study of the Urca cooling in neutron star crusts and oceans and demonstrated the necessity of performing modern shell model calculations for Gamow-Teller transition rates to understand the (anti)neutron production.

The basic structure of the PSM has already been discussed in detail in our previous article \cite{jiaqi2023}. Therefore, here, for the present study with odd-$A$ nuclei, we only mention the general differences in the selection of parameters and construction of configuration space from our previous study where even-even systems were treated. The Hamiltonian adopted in the PSM contains separable forces:
\begin{equation}\label{two-body}
\hat{H}=\hat{H_{0}}-\frac{1}{2}\chi_{QQ}\sum_\mu \hat{Q}_{2\mu}^{\dagger} \hat{Q}_{2\mu}-G_{M}\hat{P}^{\dagger}\hat{P}-G_{Q}\sum_{\mu}\hat{P}_{2\mu}^{\dagger}\hat{P}_{2\mu},
\end{equation}
where $\hat{H_{0}}$ is the SP term which includes the spin-orbit force. 
The remaining terms in Eq.~(\ref{two-body}) represent two-body interactions, namely, the quadrupole-quadrupole, monopole-pairing, and quadrupole-pairing interactions, respectively. The quadrupole-quadrupole interaction strength $\chi_{QQ}$ is determined in a self-consistent way such that the input quadrupole deformation and the one resulting from the HFB procedure coincide with each other \cite{PSM_Hara_Sun}. The pairing forces in Eq.~(\ref{two-body}) are assumed to be of the isovector type, i.e., they act only between the like nucleons. The coupling constant for the monopole-pairing force is taken to be of the following form:
\begin{equation}
G _{M} = \left(G_{1} \mp G_{2} \frac{N - Z}{A}\right)\frac{1}{A} ~~~~~~\textnormal {(in MeV)},
\end{equation}
where, $``+" ~(``-")$ denotes protons (neutrons). The choice of the monopole pairing strengths $G_1$ and $G_2$ depends on the size of the SP space in the calculation, and their values are adjusted to reproduce the experimental odd-even mass differences in the respective nuclear mass region. The strengths $G_1$ and $G_2$ are fixed as $20.12$ and $13.13$, respectively, for the present study. Note that there is a $\sim 5\%$ damping in the values of these two parameters compared to what we considered in our previous study for the even-even case of $^{164}$Dy  ($G_1=21.24$ and $G_2=13.86$) \cite{jiaqi2023}. This is to account for the weakened pairing correlation in the odd-$A$ systems relative to the neighboring even-even ones (see discussions below). Lastly, the quadrupole pairing strength $G_Q$ is assumed to be proportional to $G_M$, and the proportionality constant is usually taken in the range of $0.16 - 0.20$ \cite{jiaqi2023}. 

While the same form of PSM Hamiltonian, as given in Eq. (\ref{two-body}), is used for all nuclei irrespective of whether they are even-even, even-odd, odd-even, or odd-odd, distinct multi-qp configuration spaces are defined for each of the four types. The multi-qp configuration spaces for odd-neutron and odd-proton nuclei, built on the deformed 1-qp basis states, are given as:
\begin{equation}\label{shell_config}
\begin{aligned}
\{a_{\nu_{i}}^{\dagger}\ket\phi, a_{\nu_{i}}^{\dagger}a_{\nu_{j}}^{\dagger}a_{\nu_{k}}^{\dagger}\ket\phi,
a_{\nu_{i}}^{\dagger}a_{{\pi}_{j}}^{\dagger}a_{{\pi}_{k}}^{\dagger}\ket\phi, a_{\nu_{i}}^{\dagger}a_{\nu_{j}}^{\dagger}a_{\nu_{k}}^{\dagger}a_{\pi_{l}}^{\dagger}a_{{\pi}_{m}}^{\dagger}\ket\phi, ~\dots\}\\\text{for odd-neutron nuclei},\\
\{a_{\pi_{i}}^{\dagger}\ket\phi, a_{\pi_{i}}^{\dagger}a_{\pi_{j}}^{\dagger}a_{\pi_{k}}^{\dagger}\ket\phi,
a_{\nu_{i}}^{\dagger}a_{{\nu}_{j}}^{\dagger}a_{{\pi}_{k}}^{\dagger}\ket\phi, a_{\nu_{i}}^{\dagger}a_{\nu_{j}}^{\dagger}a_{\pi_{k}}^{\dagger}a_{\pi_{l}}^{\dagger}a_{{\pi}_{m}}^{\dagger}\ket\phi, ~\dots\}\\\text{for odd-proton nuclei}.
\end{aligned}
\end{equation}

Once the configuration space is determined, the rest of the calculations are the same as in the even-even nuclei \cite{jiaqi2023}. The set of qp states in Eq.~(\ref{shell_config}) is transferred to the laboratory frame by the angular momentum projection technique which restores the rotational symmetry initially violated in the deformed Nilsson basis. The shell model diagonalization is then carried out in the laboratory frame which generates {\it eigenstates of angular momentum and parity}.

\section{RESULTS AND DISCUSSION}\label{PSM_results}

There is one single neutron (proton) in the ground state in the odd-neutron (odd-proton) nuclei which is blocked from pair formation. This feature plays a conclusive role in the behavior of odd-$A$ NLDs. The ground state and low-lying energy states in an odd-$A$ nucleus consist of 1-qp configurations (instead of the merely 0-qp state, i.e. the qp vacuum state in even-even nuclei). Usually, a few (and maybe up to ten) deformed 1-qp states exist around the Fermi levels in odd-$A$ nuclei. All rotational and vibrational types of collective excitations as well as multi-qp excitations are based on these 1-qp configurations. Therefore, at equal conditions, NLDs in odd-$A$ nuclei at a given excitation are, on average, a few to ten times higher than in their neighboring even-even adjoins. 

Moreover, as we will see in later discussions, not only does the single unpaired nucleon in an odd-$A$ nucleus influence the NLD curve plotted as the function of excitation energy, but also significantly influences the parity- and the spin-dependent NLDs. Pairing, which is viewed as one of the most important correlations that determine many basic properties of atomic nuclei, also governs the NLD properties of odd-$A$ systems. As we will demonstrate below, at a given excitation energy, higher density of levels in odd-neutron $^{163}$Dy and odd-proton $^{163}$Tb than in the adjacent even-even $^{164}$Dy is a direct consequence of weaker pairing correlations in the odd-$A$ nuclear systems. 

\begin{figure}\centering
\includegraphics[width=3.8in]{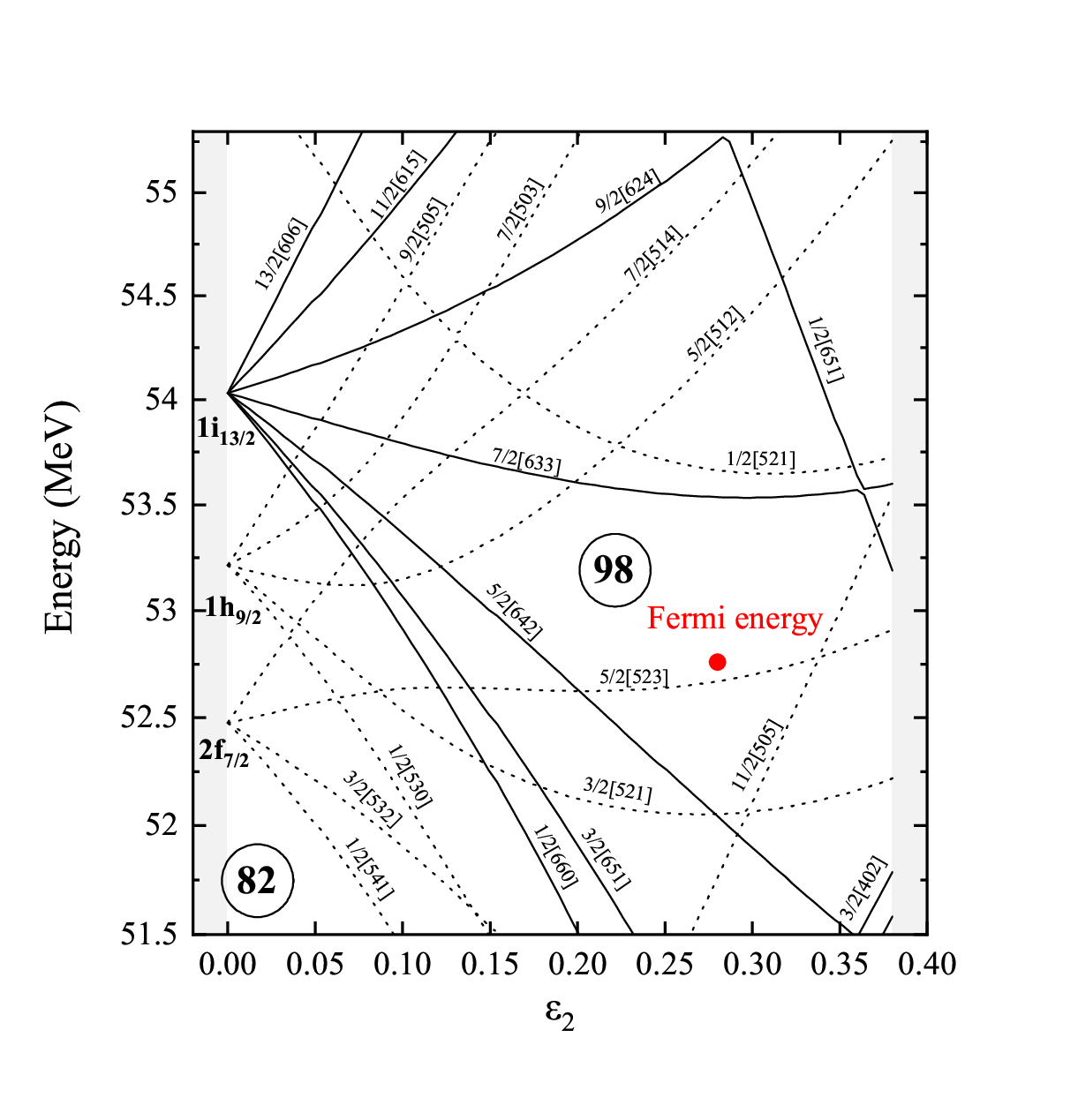}
\caption{\label{Nilsson-Dy163}(Color online) Neutron Nilsson diagram for deformed rare-earth nuclei. The Fermi energy in $^{163}$Dy corresponding to the deformation $\varepsilon_2$ = 0.28 is marked with a red dot.}
\end{figure}

\begin{figure}\centering
\includegraphics[width=3.8in]{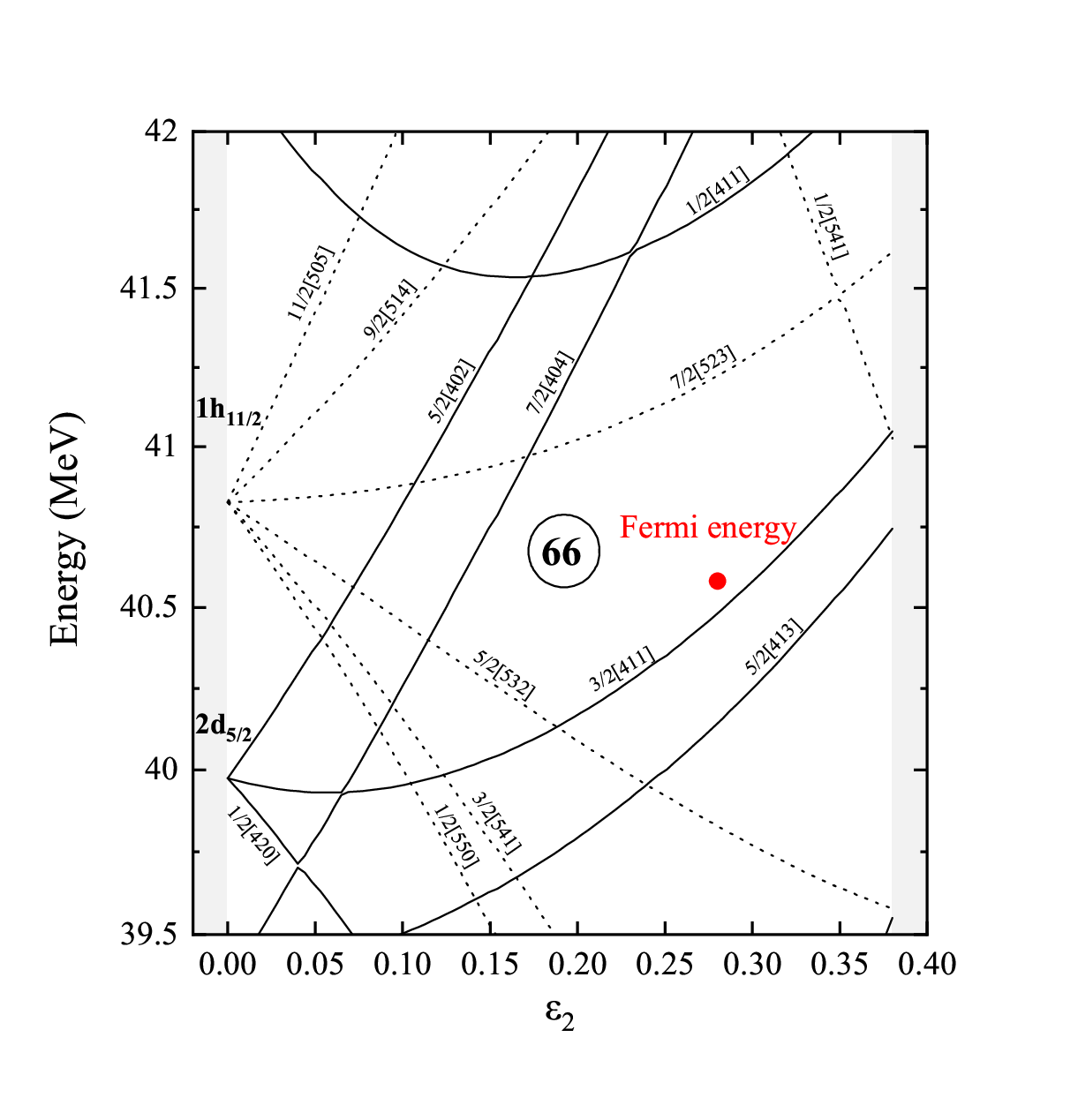}
\caption{\label{Nilsson-Tb163}(Color online) Proton Nilsson diagram for deformed rare-earth nuclei. The Fermi energy of $^{163}$Tb corresponding to the deformation $\varepsilon_2$ = 0.28 is marked with a red dot.}
\end{figure}

\begin{figure*}
\includegraphics[width=6.4in]{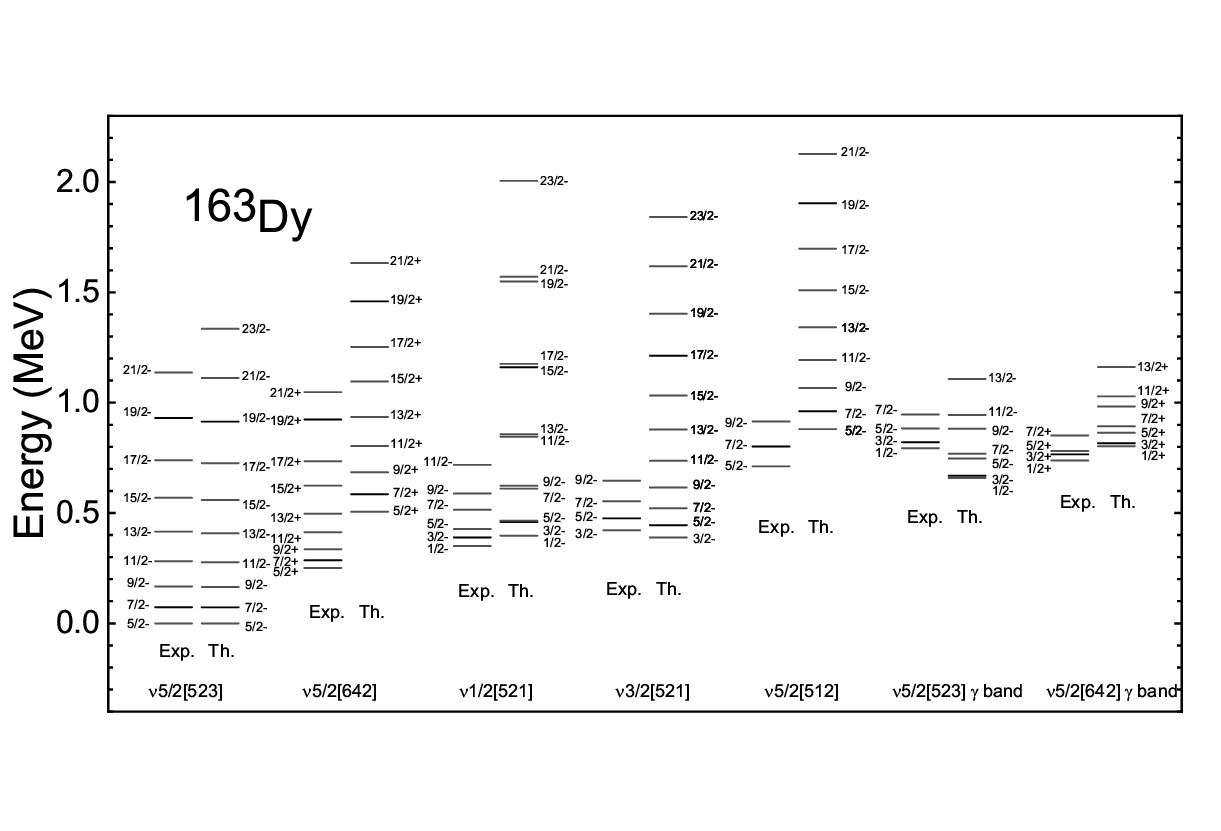}\caption{\label{Dy163-energylevels} Energy levels belonging to the low-lying rotational bands in $^{163}$Dy from the present calculation are compared with discrete levels of known spin and parity obtained via spectroscopic measurements and available in the NNDC database \cite{NNDC}.}
 \end{figure*}

\begin{figure*}  \includegraphics[width=6.4in]{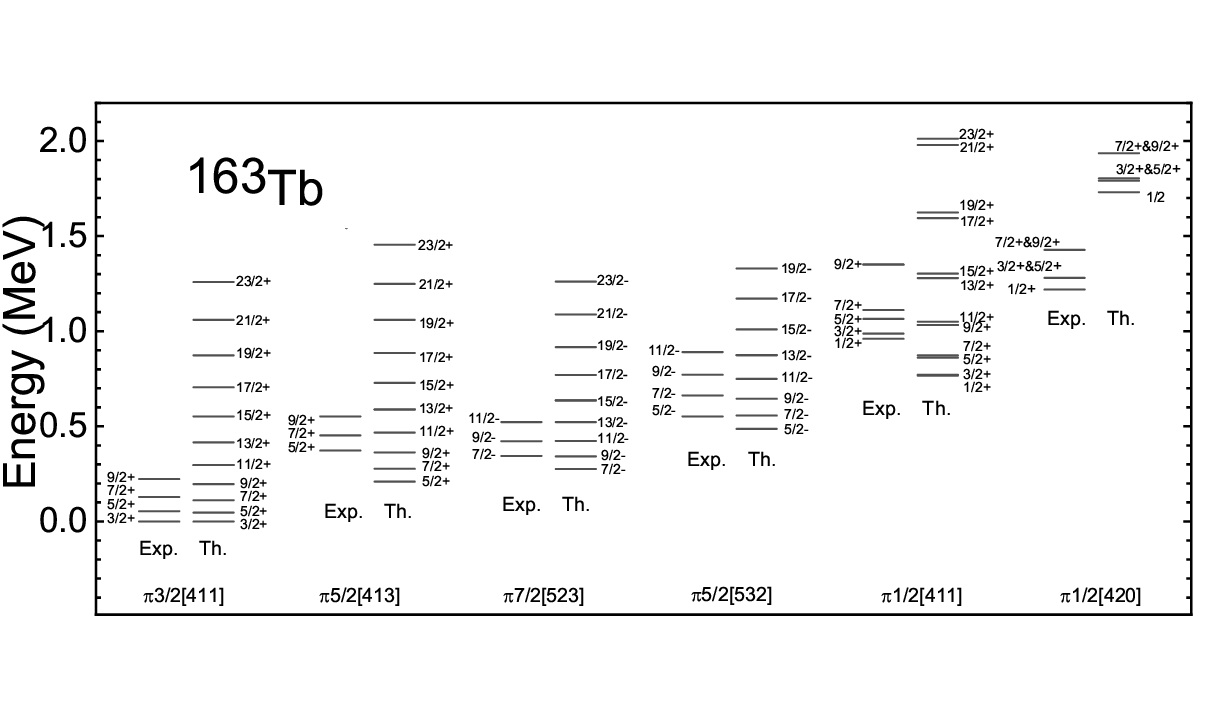}\caption{\label{Tb163-energylevels} Energy levels belonging to the low-lying rotational bands in $^{163}$Tb from the present calculation are compared with discrete levels of known spin and parity obtained via spectroscopic measurements and available in the NNDC database \cite{NNDC}.}
\end{figure*}

\subsection{Distribution of deformed SP states in the Nilsson diagram}\label{SP_distribution}

For well-deformed even-even nuclei, such as $^{164}$Dy in our preceding study \cite{jiaqi2023}, the structure of the low-energy NLD below 1 MeV is rather simple. The states there all belong to collective excitations among which, there are $I^{\pi} = 2^+, 4^+$, and $6^+$ states of the ground-state rotational band and a few collective vibrational states. However, the structure of low-lying levels in the odd-$A$ nuclei is generally more complicated. In odd-$A$ nuclei, all low-lying states have the structure of unpaired single-qp states. These 1-qp states, being single-neutron (single-proton) states in even-odd (odd-even) nuclei (see Eqs. (\ref{shell_config})), are characterized by their respective intrinsic quantum numbers in the deformed potential. Figures~\ref{Nilsson-Dy163} and \ref{Nilsson-Tb163} illustrate the distribution of deformed single-neutron states in $^{163}$Dy and single-proton states in $^{163}$Tb in the form of Nilsson diagrams, which are generated by using the standard Nilsson parameters, $\kappa$ and $\mu$, in Ref.~\cite{Nil-1985}.

The Nilsson diagram is a widely used tool for discussing the structure of experimental nuclear states. Especially for axially-deformed odd-$A$ nuclei, the SP orbitals in Nilsson diagrams usually have a one-to-one correspondence to the band-head configurations of experimentally observed rotational bands. As we shall demonstrate later in the paper, the Nilsson SP distribution in the vicinity of the Fermi level (denoted as Fermi energy in Figs.~\ref{Nilsson-Dy163} and \ref{Nilsson-Tb163}) can provide an estimate for the energy shift between NLDs \cite{Guttormsen2000} in odd-$A$ nuclei and their even-even adjoins. Using the Figs. \ref{Nilsson-Dy163} and \ref{Nilsson-Tb163}, one can estimate the characteristic SP level density ($d_\Delta$), by counting all the deformed SP Nilsson orbitals whose quasiparticle energies fall within the energy interval of one $\Delta$ above the Fermi level and one $\Delta$ below the Fermi level, where $\Delta$ is the BCS pairing gap energy. A glance at Figs. \ref{Nilsson-Dy163} and \ref{Nilsson-Tb163} can immediately tell us that $d_\Delta$ is a varying quantity for varying nuclei. It can be considerably different for different nuclei depending on where the Fermi energy ($E_{\rm Fermi}$) lies in the Nilsson diagram. When $E_{\rm Fermi}$ lies in a dense region of deformed orbitals, $d_\Delta$ is large and so is the NLD and vice versa. This fact depicts how the microscopic structure information is linked with the global behavior of odd-$A$ NLDs. Throughout the paper, we shall frequently refer to Figs.~\ref{Nilsson-Dy163} and \ref{Nilsson-Tb163} whenever the discussion is relevant.

\subsection{Analysis of discrete levels in $^{163}$Dy and $^{163}$Tb}\label{discrete_level}

Though the available information of experimentally observed discrete nuclear levels in the open database (e.g., in the NNDC database \cite{NNDC}), is far from being regarded complete, they are essential for benchmarking the NLD model calculations. Without a quantitative comparison with these known levels, one may doubt the model’s predictive ability for cases for which data do not exist. The Oslo method, which is one of the most advanced experimental methods for the determination of NLD in the quasi-continuum region of excitation, uses the known discrete levels at the low-energy anchor region for normalization to determine the slope of the Oslo NLD curve \cite{oslo_Midtbo}.

In Fig.~\ref{Dy163-energylevels}, we show energy levels belonging to the low-lying rotational bands in the even-odd nucleus $^{163}$Dy from the present calculation and compare them with the discrete level data with known spin and parity available in the NNDC database \cite{NNDC}. From Fig. \ref{Nilsson-Dy163}, one can find that the nearest Nilsson orbits to the neutron Fermi energy level, $E_{\rm Fermi}$, are the odd-parity $\nu 5/2^-[523]$ and even-parity $\nu 5/2^+[642]$, which correspond to, respectively, the band-head configuration of the ground-state band and the lowest excited band shown in Fig.~\ref{Dy163-energylevels}. There are other odd-parity orbits, e.g., $\nu 1/2^-[521]$ above $E_{\rm Fermi}$ and $\nu 3/2^-[521]$ below $E_{\rm Fermi}$. All these Nilsson orbits lie roughly within 1 MeV (or more precisely, the rough measure of the BCS pairing gap $\Delta$ as mentioned earlier) above and below $E_{\rm Fermi}$ at the deformation $\varepsilon_2=0.28$ (see Fig. \ref{Nilsson-Dy163}). They are the lowest 1-qp configurations in $^{163}$Dy, having a one-to-one correspondence with the known discrete level data \cite{NNDC}. Close to $E_{\rm Fermi}$, there are two other orbits, even-parity $\nu 7/2^+[633]$ and odd-parity $\nu 11/2^-[505]$. As we shall see below, the structure of these 1-qp states plays important roles in determining the property of NLD in $^{163}$Dy. From the above discussion, it is obvious that in $^{163}$Dy, at low excitations, a non-equal distribution of odd- and even-parity levels is expected with the odd-parity levels being more in number relative to the even-parity ones. 

Below 1 MeV of excitation, discrete nuclear levels in $^{163}$Dy include two $\gamma$ vibrational bands, which can be interpreted as the configurations formed from the coupling of a $2^+$ $\gamma$ phonon to the lowest two 1-qp states of this nucleus. Between the two right-most $\gamma$ bands shown in Fig. \ref{Dy163-energylevels}, the one with bandhead spin-parity $1/2^-$ has the structure of the 1-qp ground state $\nu 5/2^-[523]$ coupled to a $2^+$ $\gamma$ phonon, while the other one with bandhead spin-parity $1/2^+$ has the structure of the excited 1-qp state $\nu 5/2^+[642]$ coupled to a $2^+$ $\gamma$ phonon.

Similarly, Fig.~\ref{Tb163-energylevels} shows energy levels for the low-lying excited bands in the odd-even nucleus $^{163}$Tb from the present calculation which have been compared with the experimental discrete level data of known spin and parity available in the NNDC database \cite{NNDC}. The isotope $^{163}$Tb is radioactive with a half-life of about 20 minutes, and hence, the known spectroscopic data are scarce. Therefore, most of the energy levels in $^{163}$Tb are from our theoretical prediction. From Fig. \ref{Nilsson-Tb163}, one can see that the nearest Nilsson orbit to the proton Fermi energy level, $E_{\rm Fermi}$, is the even-parity $\pi 3/2^+[411]$, which corresponds to the band-head configuration of the ground-state. Below it, there is an even-parity orbit $\pi 5/2^+[413]$. There are two other odd-parity orbits: $\pi 7/2^-[523]$ above $E_{\rm Fermi}$ and $\pi 5/2^-[532]$ below $E_{\rm Fermi}$. It can be realized from Fig.~\ref{Tb163-energylevels} that these orbitals correspond to the lowest three 1-qp configurations of excited rotational bands. Other energy levels in $^{163}$Tb are predicted to lie high in energy, typically at about or higher than 1 MeV of excitation. The two right-most 1-qp bands in Fig.~\ref{Tb163-energylevels} contain examples of such levels. Our calculation correctly describes the characteristic degeneracy feature of the levels of these two excited even-parity 1-qp bands although it does not reproduce their excitation energies quite well. By degeneracy feature we mean, among the two right-most bands in Fig.~\ref{Tb163-energylevels}, $1/2^+$ and $3/2^+$ states, $5/2^+$ and $7/2^+$ states, etc., in the former $\pi 1/2^+[411]$ band have nearly degenerate energies whereas $3/2^+$ and $5/2^+$ states, $7/2^+$ and $9/2^+$ states, etc., in the latter $\pi 1/2^+[420]$ band are perfectly degenerate. The physical mechanism leading to such degeneracy was discussed in Ref.~\cite{Sun1994b}.

\subsection{Structure-dominant NLDs in $^{163}$Dy and $^{163}$Tb}\label{NLDs}

\begin{figure}
\includegraphics[width=3.50in]{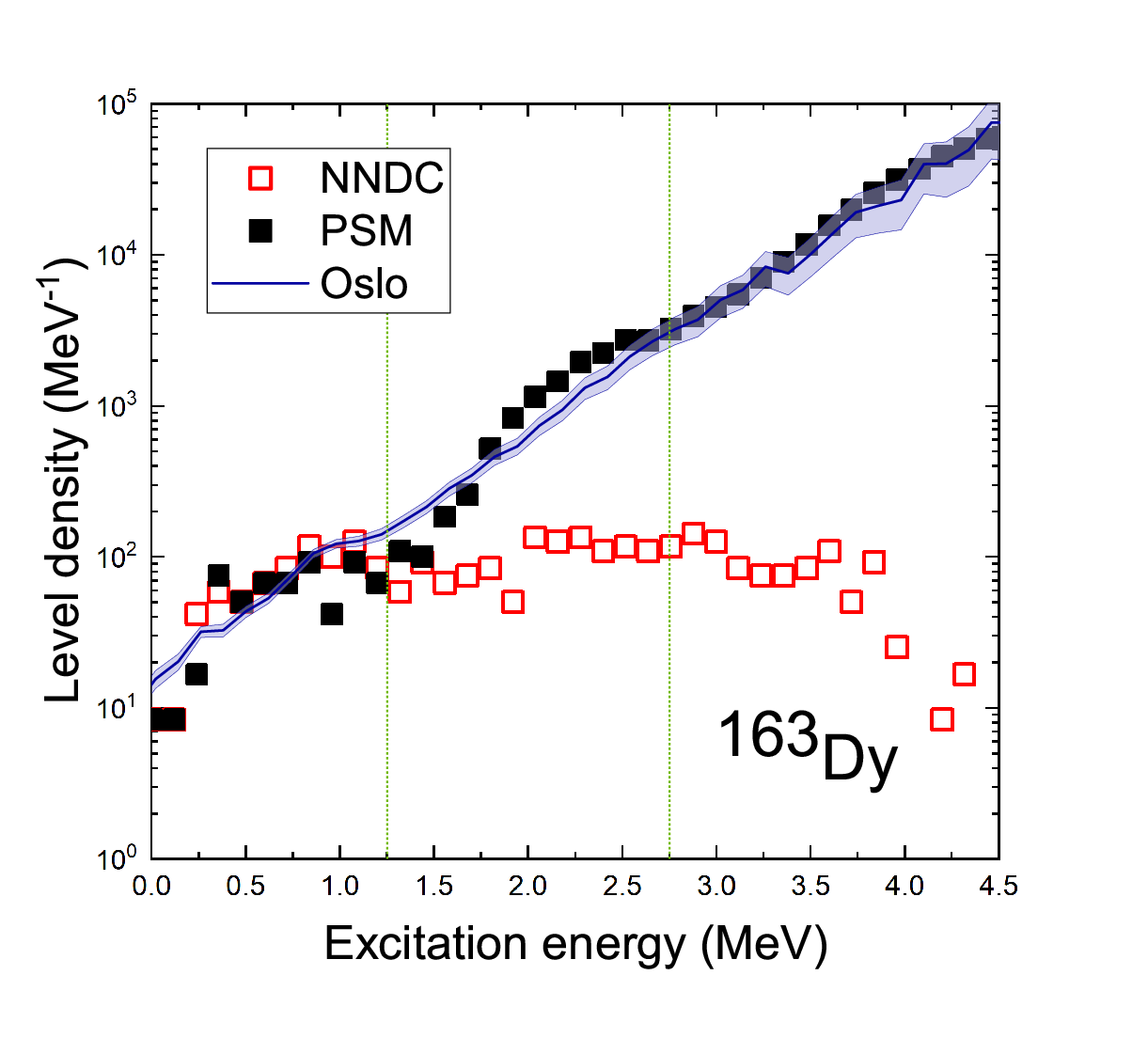} 
\caption{\label{Dy163-major}(Color online) Level densities (filled squares) in the excitation energy bins of $^{163}$Dy, calculated in our projected shell model method, are compared with the ones derived from the spectroscopically measured discrete level data (red open squares) \cite{NNDC} and the level density data provided by the Oslo method (solid blue line, shaded area representing uncertainty) \cite{Oslo-Dy163}.}
\end{figure}

\begin{figure} \includegraphics[scale=0.47]{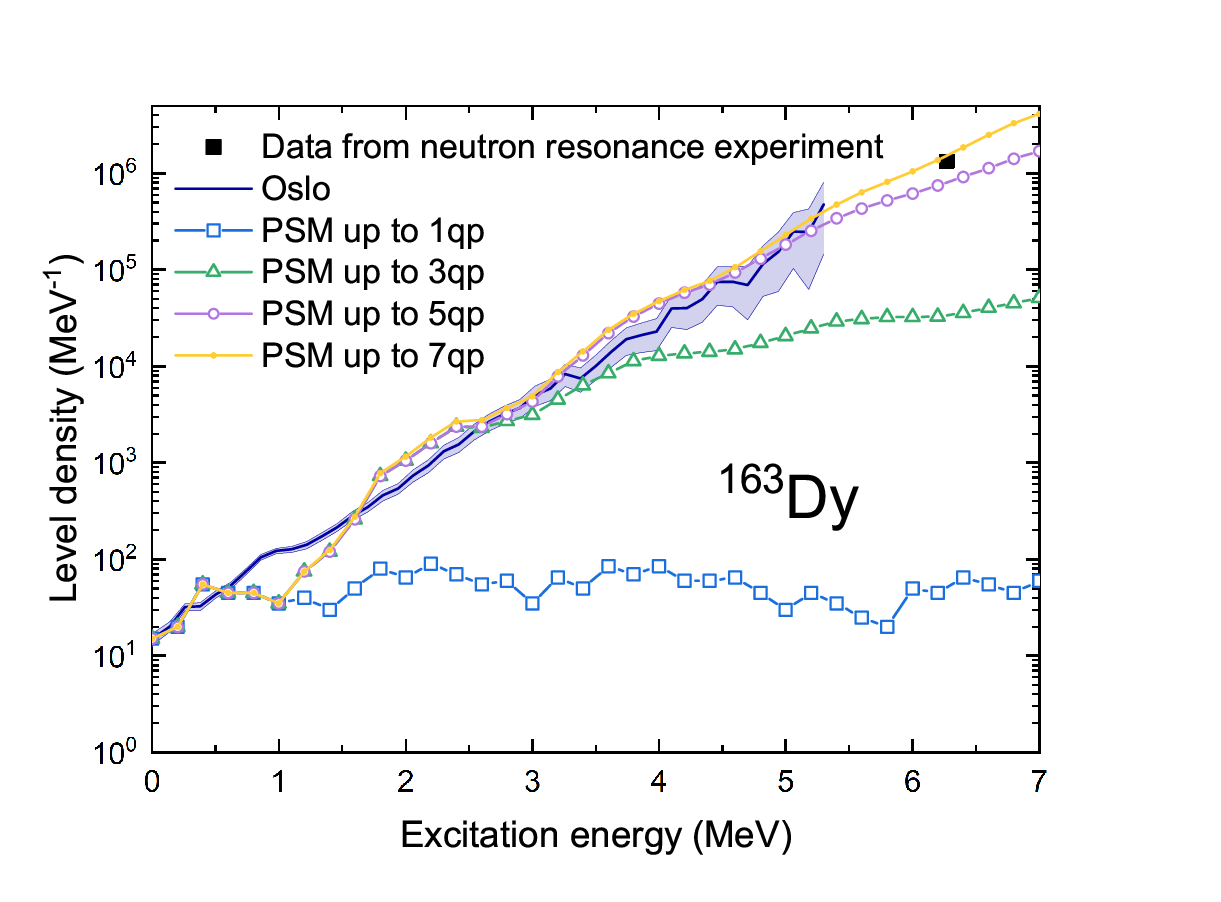} \caption{\label{Dy163-ldbeforediag}(Color online) Unmixed NLDs (see text for definition) in the excitation energy bins of even-odd nucleus $^{163}$Dy, resulting from the contribution of 1-qp, 3-qp, 5-qp, and 7-qp configurations are shown along with the Oslo NLD data (solid blue line, shaded area representing uncertainty) \cite{Oslo-Dy163} and the NLD value at the neutron separation energy obtained using the measured neutron resonance spacing data (filled black square).}
\end{figure}

\begin{figure} \includegraphics[width=3.50in]{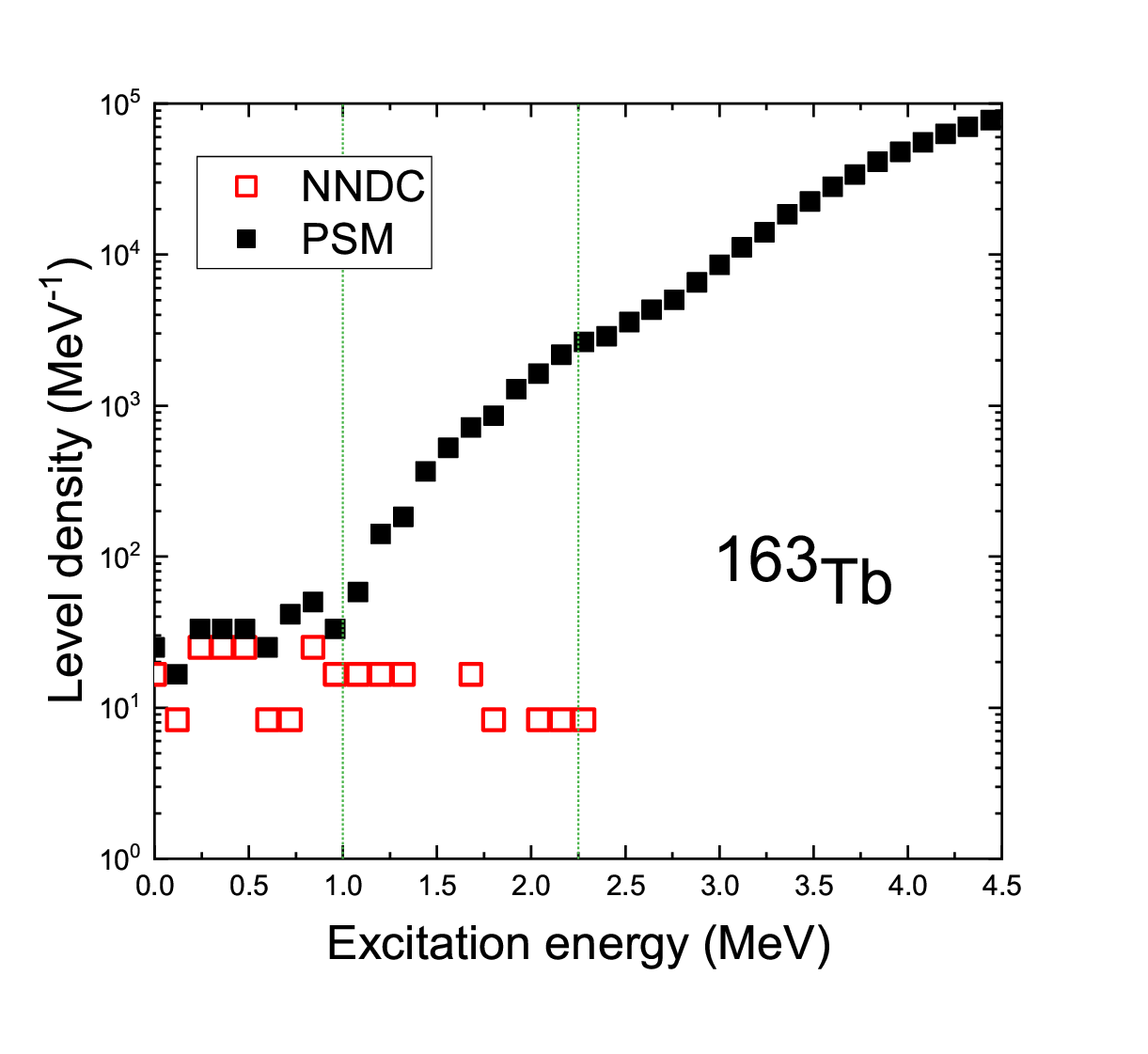} 
\caption{\label{Tb163-major}(Color online) Level density (filled squares) in excitation energy bins of $^{163}$Tb, calculated in our projected shell model method, is compared with the ones derived from the spectroscopically measured discrete level data (red open squares) \cite{NNDC}.}
\end{figure}

\begin{figure} \includegraphics[scale=0.47]{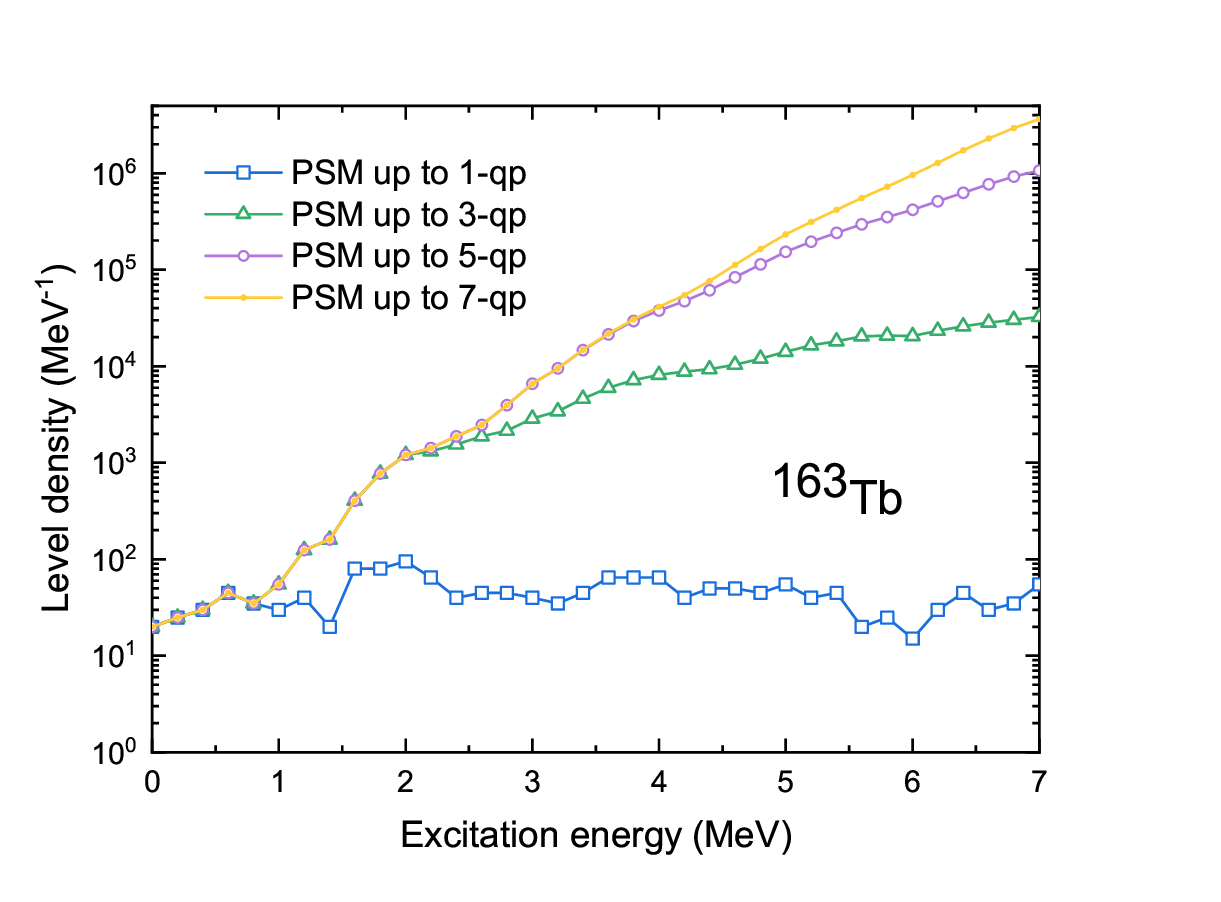}  \caption{\label{Tb163-ldbeforediag}(Color online) Unmixed NLDs (see text for definition) in the excitation energy bins of odd-even nucleus $^{163}$Tb, resulting from the contribution of 1-qp, 3-qp, 5-qp, and 7-qp configurations.}
\end{figure}

With projected shell model calculations, the obtained NLDs, $\rho (E, I, \pi)$, are explicit functions of excitation energy, spin, and parity for which, they can be directly supplied to reaction calculations. On the other hand, as discussed in Section \ref{intro}, experimental NLDs and most of the calculated NLDs by other theories do not have definite spin and parity. To compare our results with those NLDs, in Figs. \ref{Dy163-major} - \ref{Tb163-ldbeforediag}, we present our calculated NLDs summed over both positive and negative parities as well as summed over all spins, defined as:
\begin{equation}\label{rho}
    \rho(E) = \sum_\pi \sum_{I\le 10} \rho (E, I, \pi).
\end{equation}
For most of the known discrete levels in the rare-earth nuclei with experimentally assigned spins, except for strongly populated collective bands, such as the ground-state band and some vibrational bands, where the spectroscopic measurement can be extended to high-spin states, spin quantum numbers seldom exceed 10$\hbar$. This entails us using a spin cut-off in the summation of Eq.~(\ref{rho}) for comparison purposes. Therefore, unless otherwise mentioned, our presented results contain the levels with spin $I \le 10\hbar$. This spin-cut-off restriction can be released if we want high-spin states to be included in NLDs. In the later part of the paper, we shall deal with $\rho (E, I, \pi)$ to discuss the spin- and the parity-distribution in NLDs. 

In Fig.~\ref{Dy163-major}, we show the calculated level density (filled squares) for $^{163}$Dy, together with the experimental one obtained from the known discrete levels (red open squares) \cite{NNDC} and the level density curve of the Oslo method (blue solid line) taken from Ref.~\cite{Oslo-Dy163}. All the NLDs are consistently presented in 0.12-MeV energy bins. As one can see, a good agreement between our calculation and experimental discrete level density is found up to 1.25 MeV in excitation, i.e., the energy up to which the measured discrete levels are believed to be complete. The Oslo NLD curve does not contain much structure details and begins to take the straight-line behavior from the energy as low as 1 MeV. In contrast, our PSM calculation suggests some step-wise variations up to 2.75 MeV, which are absent in the Oslo curve. At 2.75 MeV, our calculated NLD starts to take a straight line, coinciding perfectly with the Oslo curve for the high-energy region. 

In the same way as in our previous publication where we presented NLD calculation in the even-even $^{164}$Dy nucleus \cite{jiaqi2023}, we divide the entire excitation energy range of Fig.~\ref{Dy163-major} representing NLD in even-odd $^{163}$Dy, into three regimes according to the level properties therein. We use two vertical green lines to separate the three regimes. We call the energy interval $0-1.25$ MeV the unpaired regime since here, the level structure is entirely due to the unpaired single neutrons. The interval $1.25-2.75$ MeV is termed as the pair-breaking regime, which corresponds to the region where the process of proton-pair-breakings commences causing prominent structure effects in NLDs (which will be discussed later in detail). 
Finally, we name the energy region $> 2.75$ MeV as the multi-qp (or chaotic) regime, marked by the beginning of the creation of 5-qp states (and later 7-qp, 9-qp, etc. states) in the system through the coupling of the unpaired neutron with multiple (two and more) broken nucleon pairs. Therefore, beyond $> 2.75$ MeV, the NLD structure is dominated by such high-order multi-qp states justifying such nomenclature. The NLD, as can be seen from Fig.~\ref{Dy163-major}, rapidly grows in number reaching a magnitude $> 10^{3}$ per MeV, indicating the evolution of the system from a single-particle state to a  chaotic state \cite{Zelevinsky2019}. As we will see in the following sections, different from the pair-breaking regime, only the chaotic region shows statistical properties in NLDs characterized by an equal division of even- and odd-parity levels and Gaussian distribution of level-spins. Here, a difference should be noted that while in our previous publication for even-even nuclei, we termed the first excitation energy regime of the NLD plot as the collective regime \cite{jiaqi2023}, in the present case for the odd-$A$ nuclei, we term the same as the unpaired regime.

Our PSM calculation for NLD with exact shell model diagonalization in Fig.~\ref{Dy163-major} is performed up to 4.5 MeV of excitation where the density of nuclear levels already reaches the order of $10^5$ per MeV. Extension of calculation to higher spin states and/or to higher excitations is possible. However, when the configuration space is enlarged further, the computation becomes time-consuming, which hinders general applications and discussions for the still higher excitations. To get around the dilemma, our theoretical NLDs can be calculated with the states we obtain just when the angular momentum projection calculation is over but the diagonalization calculation is yet to start. In other words, if we entirely disregard the final step of matrix diagonalization, which is meant to attain the mixing of configurations, with a decent computation cost, we can obtain levels that are qualitatively similar to those obtained by full PSM calculation. This is because, among the several steps involved in a complete PSM calculation, the last step of Hamiltonian matrix diagonalization is the one that renders the huge computation-time cost when the configuration space is expanded. The so-obtained nuclear levels without performing the matrix diagonalization correspond to those of the diagonal elements with which the mixing of off-diagonal elements is neglected. Note that such levels also preserve spin and parity as good quantum numbers. Hereafter, we call the calculated NLDs using these levels as `unmixed' NLDs. 

Thus, with a much-reduced computation effort, we can easily extend our NLD calculation beyond the neutron separation energy. An additional advantage with unmixed nuclear levels is that we can separately plot NLDs built from specific orders of qp configurations. In this way, the dominance of different orders of qp configurations over distinct excitation energy regions in the total NLD can be clearly identified. In Fig.~\ref{Dy163-ldbeforediag}, we present unmixed NLDs in $^{163}$Dy from separate orders of qp configurations. For example, the one labeled as `PSM up to 7qp' represents the density of unmixed levels calculated using 
all 1-qp, 3-qp, 5-qp, and 7-qp states.  

Several noticeable consequences can be discussed using Fig.~\ref{Dy163-ldbeforediag}. First, this plot may answer the basic question of how multi-qp configurations contribute to the exponential increase of the total level density with excitation energy. Recall that our PSM adopts three neutron harmonic oscillation shells ($N=4, 5$, and $6$) in the model space for nuclei of the rare-earth sector, which can allow a maximal number of 64 deformed single-neutrons to occupy. Within this model space, the calculated NLD with the lowest order of qp configurations, i.e. the unpaired 1-qp states, can never exceed the order of $10^2$ per MeV (see open blue squares in Fig.~\ref{Dy163-ldbeforediag}).  With increasing excitation, higher-order qp states must be built into the configuration space successively to keep the NLD rising exponentially. Fig.~\ref{Dy163-ldbeforediag} depicts clear step-like structures for multi-qp contributions; or in other words, successive steps can be seen over (i) $0-1$ MeV, (ii) $1.25-2.5$ MeV, and (iii) $2.75-4.35$ MeV bins due to dominant contributions from (i) 1-qp states, (ii) 3-qp states, and (iii) 5-qp states, respectively. For energies beyond $4.5$ MeV and up to the neutron separation energy, 7-qp configurations dominate. 

It is evident that up to 4.5 MeV, Fig.~\ref{Dy163-ldbeforediag} representing unmixed NLD shows quantitatively the same results as that obtained by full diagonalization calculation in Fig.~\ref{Dy163-major}. This may suggest that our angular momentum projected states are physically similar to the realistic levels obtained from the complete PSM calculation, i.e., the configuration mixing influences the final results very little. In addition, our unmixed NLD in the chaotic regime follows the experimental Oslo curve very well up to the energy at about 5.3 MeV beyond which the Oslo measurement is no longer feasible. Remarkably, further up in energy, the unmixed NLD correctly reproduces the data point determined by the neutron resonance experiment at 6.271 MeV \cite{Oslo-Dy163} where the level density value reaches more than $10^{6}$ per MeV.

In Ref.~\cite{Guttormsen2000b}, Guttormsen {\it et al.} made an interesting observation from the $^{162}$Dy and $^{172}$Yb level densities obtained in the Oslo experiment using $(^3{\rm He}, \alpha\gamma)$ reaction. By deducing entropy within the microcanonical ensemble from the extracted Oslo NLD data for the two pairs of adjacent odd-$A$ and even-even deformed heavy nuclei in the rare-earth sector, the authors made a comparative analysis and gave an interesting conclusion in terms of quasiparticle excitations \cite{Guttormsen2000b}. They extracted the number of excited quasiparticles created over the energy range to indicate the phenomena of successive orders of qp state formation due to the breaking of the nucleon pairs. They concluded that at an excitation energy of $\sim$ 5.5 MeV in these two isotopes, the maximum number of excited quasiparticles is $\sim 6$, corresponding to the breaking of three nucleon pairs. This experimental finding from the even-even isotopes is consistent with our present results in Fig.~\ref{Dy163-ldbeforediag} for the even-odd $^{163}$Dy, where, from $\sim$ 4.5 MeV and up (in the multi-qp regime), dominating contribution from 7-qp states corresponds to breaking of three pairs leading to creation of six excited quasiparticles with which the 1-qp basis states couple to form multitude of 7-qp levels. 

A similar discussion for the odd-even nucleus $^{163}$Tb is given using Fig.~\ref{Tb163-major} and Fig.~\ref{Tb163-ldbeforediag}. The total NLD of Eq.~(\ref{rho}) obtained by full PSM calculation including the final step of diagonalization for configuration mixing is shown in Fig.~\ref{Tb163-major} up to an excitation energy of 4.5 MeV which is compared with the NLD calculated using the available discrete level data from spectroscopic measurement. To maintain consistency, the experimental discrete level density as well as the PSM NLD are calculated using 0.12 MeV energy bins. As can be seen, despite the meager amount of complete discrete-level data, the agreement is reasonable. No Oslo NLD is available for this isotope to make a comparison which further indicates the importance of our model predictions in connection with their usefulness for the Hauser-Feshbach calculations. As can be seen, the NLD curve of odd-even $^{163}$Tb also shows step-like structures over the excitations similar to even-odd $^{163}$Dy. Fig.~\ref{Tb163-ldbeforediag} shows the unmixed NLD in $^{163}$Tb and step structures resulting from the contribution of the rising order of qp configurations. The steps formed from the dominance of 1-qp, 3-qp, and 5-qp configurations range over respectively, $0-0.8$ MeV, $1-2$ MeV, and $2.25-4$ MeV. Beyond 4 MeV, 7-qp states keep the NLD curve rising. Same as we have seen in the even-odd $^{163}$Dy, for odd-even $^{163}$Tb as well, the unmixed NLDs of Fig.~\ref{Tb163-ldbeforediag} are qualitatively similar to that of Fig.~\ref{Tb163-major} obtained with full configuration mixing calculation. Accordingly, in the same way as we did in case of $^{164}$Dy and $^{163}$Dy, the entire excitation energy range in Fig.~\ref{Tb163-major} for $^{163}$Tb is divided into three zones separated by the green vertical lines: the region from $0 - 1$ MeV is termed as unpaired regime, the next one from $1-2.25$ MeV is termed as pair-breaking regime, the third one from $2.25-4.5$ MeV is termed as the multi-qp (or chaotic) regime. A noticeable difference among Fig.~\ref{Dy163-major} for even-odd $^{163}$Dy, Fig~\ref{Tb163-major} for odd-even $^{163}$Tb, and Fig. 2 of Ref.~\cite{jiaqi2023} for even-even $^{164}$Dy is that the green vertical lines are shifted toward lower energies in both odd-$A$ nuclei relative to those in even-even $^{164}$Dy, meaning that in odd-$A$ nuclei, the pair-breaking process is completed earlier at lower energies. To give a quantitative measure, the energy that marks the beginning of the chaotic regime, signifying the region from where the NLD curve starts to attain a statistical character as the system enters a state of chaoticity, is 4.0 MeV for the even-even $^{164}$Dy (Fig. 2 in Ref.~\cite{jiaqi2023}), 2.75 MeV for even-odd $^{163}$Dy (Fig.~\ref{Dy163-major}), and 2.25 MeV for odd-even $^{163}$Tb (Fig.~\ref{Tb163-major}). As we shall see, this pronounced difference impact significantly the parity- and the spin-distribution in levels within odd-$A$ nuclei.

\subsection{Parity-distribution in NLDs of odd-$A$ $^{163}$Dy and $^{163}$Tb relative to that of their even-even adjoin $^{164}$Dy}\label{parity}

In addition to the excitation energy, nuclear level densities are also functions of spin and parity. The models based on simple statistics that do not consider any structure information usually assume that all the nuclear states, thus NLD, can be divided into equal groups of even and odd parity \cite{Ericson1960, Huizenga1972}. However, this equal-parity assumption is not expected to be valid in the low-energy region due to the nuclear structure effects. In an even-even nucleus, typically, the energy levels consist of the ground-state band and the two common types of collective vibrational bands (i.e., $\beta$- and $\gamma$-vibrational bands), which are all of even parity. Odd-parity states occur in the low excitations only if the nucleus has collective octupole vibrational bands. Therefore, in even-even nuclei, even-parity states are always dominant in the collective NLD regime.

\begin{figure}
\includegraphics[width=3.0in]{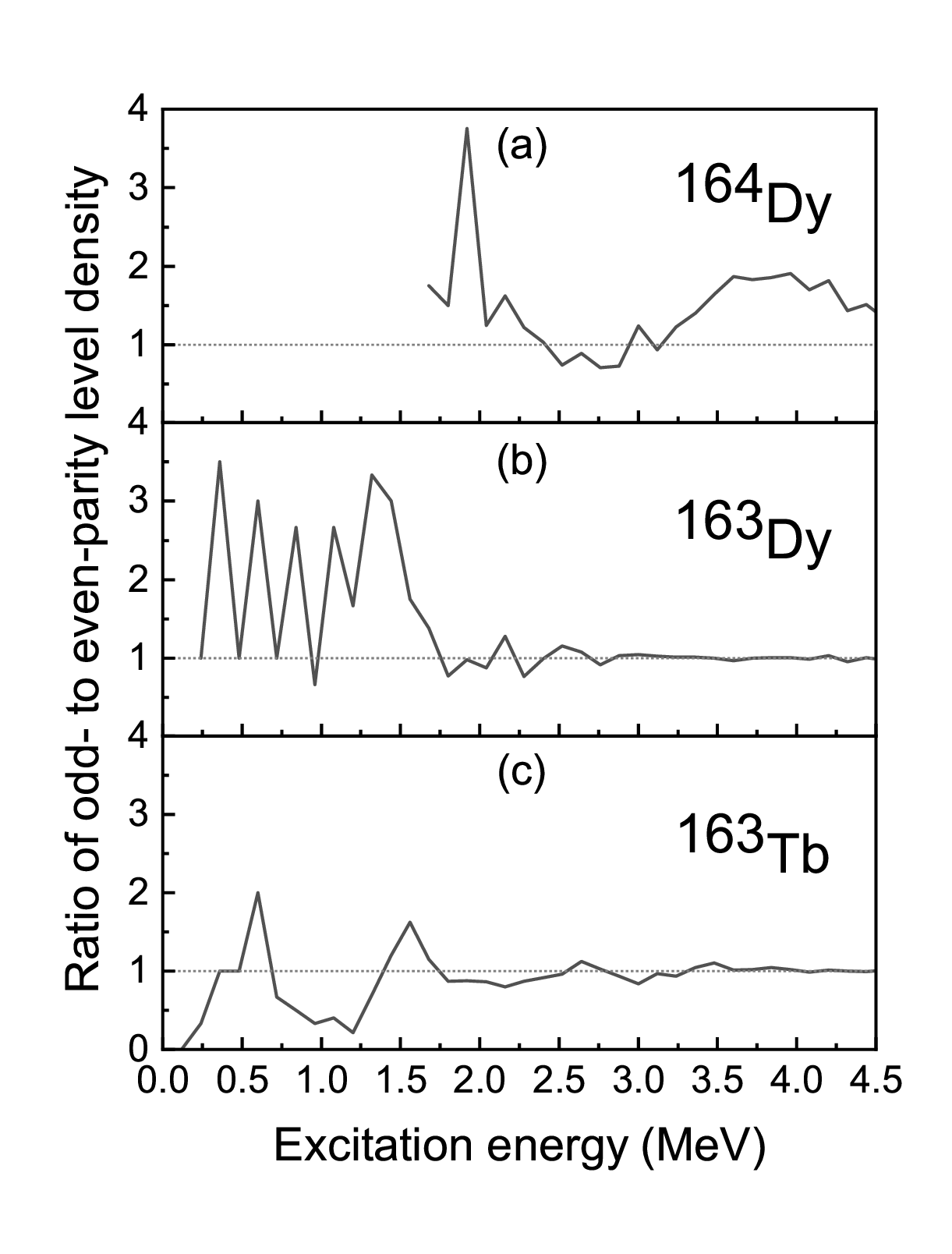}\caption{\label{ratiocompare}Ratio of odd- and even-parity level density in the excitation energy bins of (a) even-even $^{164}$Dy, (b) even-odd $^{163}$Dy, and (c) odd-even $^{163}$Tb.}
\end{figure}

The situation is different in odd-$A$ nuclei, where the lowest excitations belong to the unpaired regime. Whether the majority of states in the unpaired regime are of even parity or odd parity depends on the structure. As we shall see below for the $^{163}$Dy and $^{163}$Tb examples, the parity dependence in the low-energy NLD in a particular odd-$A$ nucleus is uniquely determined by the parity of unpaired single-particles close to the Fermi energy of the respective nucleus. To see the parity-dependence in NLDs, it is convenient to plot the ratio of NLDs with the opposite parities at a given excitation energy, i.e., $\rho(E,-) / \rho(E,+)$, with $\rho(E,\pi)$ defined as:
\begin{equation}\label{rho1}
    \rho(E,\pi) = \sum_{I\le 10} \rho (E, I, \pi),~~~~~~(\pi = +, -), 
\end{equation}
which includes all the calculated spin states up to $I$ = 10$\hbar$. We have shown such ratios for respectively even-even $^{164}$Dy, even-odd $^{163}$Dy, and odd-even $^{163}$Tb in Fig. \ref{ratiocompare}, in three partitions: (a), (b), (c) for easy comparison.
As one can see, the ratio patterns are distinctively different for the three nuclei implying different parity dependence in the immediate neighboring nuclei. In Fig.~\ref{ratiocompare}(a) for even-even $^{164}$Dy, there is no odd-parity level below 1.5 MeV in the collective regime. For the odd-neutron nucleus $^{163}$Dy (Fig. \ref{ratiocompare}(b)), since all the bands at the low excitations within the unpaired regime are of odd parity except $5/2^+[642]$ (cf. Fig.~\ref{Dy163-energylevels}), the ratio below 1.75 MeV is on average about 2. In contrast, for the odd-proton nucleus $^{163}$Tb (Fig. \ref{ratiocompare}(c)), the ratio oscillates around unity in the low energy region, which implies that in this nucleus, there is on average an equal number of even- and odd-parity unpaired single nucleons near its Fermi energy level (cf. Fig. \ref{Tb163-energylevels}). 

A striking feature in the odd-$A$ NLDs is that going up in excitation beyond 1.75 MeV, the ratio of the odd- and even-parity NLDs approaches the unity line, as can be seen in the even-odd $^{163}$Dy (Fig.~\ref{ratiocompare}(b)) as well as in the odd-even $^{163}$Tb (Fig.~\ref{ratiocompare}(c)). This is in sharp contrast to the even-even $^{164}$Dy (Fig. \ref{ratiocompare}(a)) where the ratio in the higher energy region fluctuates away from the unity line strongly. Our calculations thus suggest that in the higher energy region beyond $\sim$ 2 MeV, the odd-$A$ nuclear systems take a half-half allocation consistent with the common statistical assumption while the even-even system still shows strong structure-dependent allocations of positive- and negative-parity states even at 4.5 MeV.

Some studies existing in the literature also reported that the basic assumption of equal-parity distribution for NLDs is often not realized. For example, in Ref.~\cite{Mocelj2005}, the authors investigated the parity-distribution for several nuclei in the Fe region and found that the equal-parity ratio was not fulfilled at low excitation energies, not even at the particle separation energies. In Ref. \cite{Alhassid2000}, from shell model Monte Carlo calculation as well as by using a simple formula, Alhassid {\it et al.} found that at low energies, only a single parity dominated and cross-over to equal-parity ratio was seen only at sufficiently higher excitation energies. Early theoretical work based on the combinatorial model \cite{Herman1987, Herman1988, Cerf1991} and equidistant model \cite{Cerf1993} also found deviations from equal-parity approximation.  

Therefore, after a close comparison between the neighboring even-even and odd-$A$ nuclear systems, our present study suggests that, whether and where the equal-parity distribution in NLDs can be realized, crucially depends on the type of nuclei. The physics behind this is that the pairing and pair-breaking phenomena act differently in the even-even systems than in the odd-$A$ systems. In the odd-$A$ nucleus, depending on whether it is even-odd or odd-even, a single nucleon of a particular kind is blocked from pair formation in the ground state, and hence, in the low-excitation region, the pair-breaking process occurs only within the other kind of nucleon-pairs. One important consequence is that odd-$A$ nuclei do not experience the process of simultaneous breaking of neutron and proton pairs. This is different from what we discussed for even-even systems in our previous study for $^{164}$Dy \cite{jiaqi2023} in which a third uprise following a plateau showed up in the NLD curve because of such phenomenon of simultaneous breaking of two different kinds of fermionic pairs. As we discussed in \cite{jiaqi2023}, the required energy for breaking a neutron- {\it and} a proton pair in an even-even system is $2(\Delta_n+\Delta_p)$, which is about 3 MeV in the $^{164}$Dy case. As many such 4-qp configurations contributed successively to the NLD, we found that the third uprise started at the 3.0-MeV bin and continued to 4.0 MeV and further up where our theoretical NLD curve began to follow the statistical behavior predicted by phenomenological models \cite{Gilbert-Cameron}. In odd-$A$ nuclei, pair breaking occurs only among one type of nucleons for which the required energy is either $2\Delta_n$ (for odd-proton nuclei) or $2\Delta_p$ (for odd-neutron nuclei), which can be estimated as $\sim$ 1.5 MeV. Beyond this energy, the pair-breaking process that causes major structure variations in NLD is essentially over. Therefore, the phenomenological suggestion of the half-half allocation of even- and odd-parity states \cite{Ericson1960, Huizenga1972} in odd-$A$ nuclei can be seen to start at much lower excitations compared to that in their even-even neighbor. 

In odd-odd systems, the ground state features one unpaired neutron as well as one unpaired proton. Therefore, the lowest configurations in odd-odd nuclei are those 2-qp states of one unpaired neutron plus one unpaired proton. Therefore, unlike in even-even and odd-$A$ nuclei, in odd-odd nuclear systems, there is no need to provide energy to break nucleon pairs at low excitations to form 2-qp states. We, therefore, expect that NLDs of the odd-odd nuclei would exhibit statistical behavior very early in excitations, naturally, earlier than what we recently found in the odd-$A$ systems. 

The conclusion that has just emerged from our present shell model calculation, that the odd-$A$ nuclear systems exhibit more pronounced statistical features than their even-even counterparts, is very gratifying and useful, which allows one to accept the equal-parity assumption for NLDs in them starting from a very low excitation energy (at least 2 MeV as we have just seen above both in $^{163}$Dy and $^{163}$Tb) and beyond. In the next subsection, we will utilize the features found in odd-$A$ nuclei to discuss the spin-distribution in detail which is a much-desired quantity for various practical applications.

\subsection{Spin-distribution in NLDs of odd-$A$ nuclei}\label{spin-distribution}

The lack of reasonable information on the spin-distribution of nuclear levels can significantly affect nuclear astrophysics calculations. Recently, Liddick {\it et al.} \cite{sensitivity_1} reported that uncertainties in NLDs from poorly constrained spin-distribution caused the cross-section predictions to vary by large factors, especially as one goes farther and farther away from the stable region. In several articles \cite{Fission_HFMC_1, Fission_HFMC_2, fission_monte-carlo, fission_monte-carlo_2}, nuclear fission was studied in the statistical Hauser-Feshbach Monte-Carlo simulations, in which each primary fission fragment was treated as a compound nucleus. In such studies, knowledge of the initial spin-distribution of primary fission fragments is crucial, since the competition between prompt neutron and $\gamma$-ray emissions following the fission sensitively depends on this distribution \cite{fission_monte-carlo}. Without this knowledge, authors had to introduce an external parameter $\alpha$ that best reproduced the average neutron multiplicity (which is a precisely measured quantity). However, it turned out to be an overestimation for the description of $\gamma$-ray spectra ~\cite{fission_monte-carlo}. Despite considerable efforts \cite{Voinov2007, Grimes2013, Grimes2016}, understanding the distribution of spins in nuclear level densities remains a theoretical challenge. 

\begin{figure}
\includegraphics[width=3.5in]{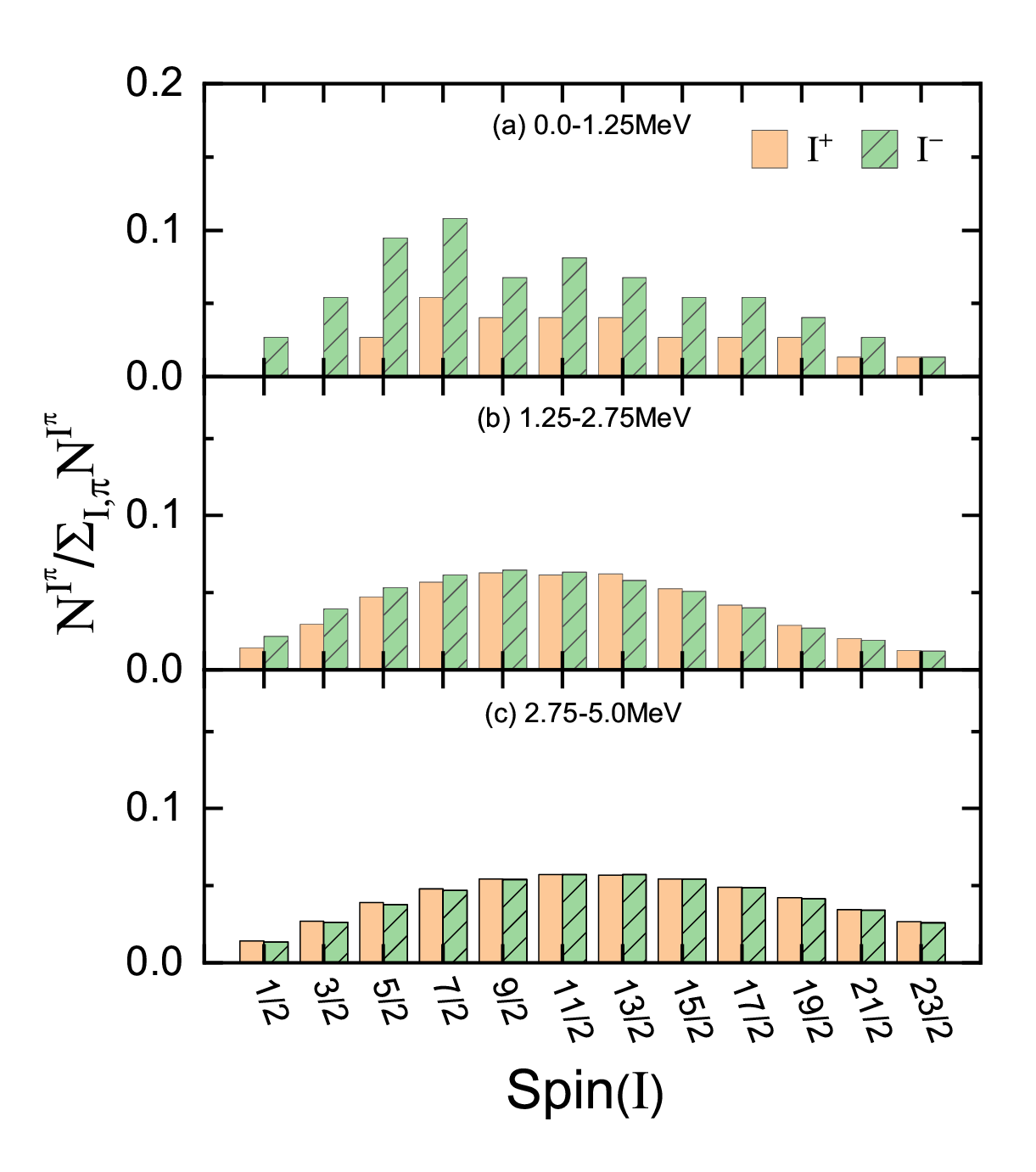} 
\caption{\label{Dy163-spin}(Color online) Histograms showing counts (normalized to unity) separately of odd-parity and even-parity nuclear levels in $^{163}$Dy sorted according to their spin values in the (a) unpaired, (b) pair-breaking, and (c) multi-qp excitation energy regimes.}
\end{figure}

\begin{figure}
\includegraphics[width=3.5in]{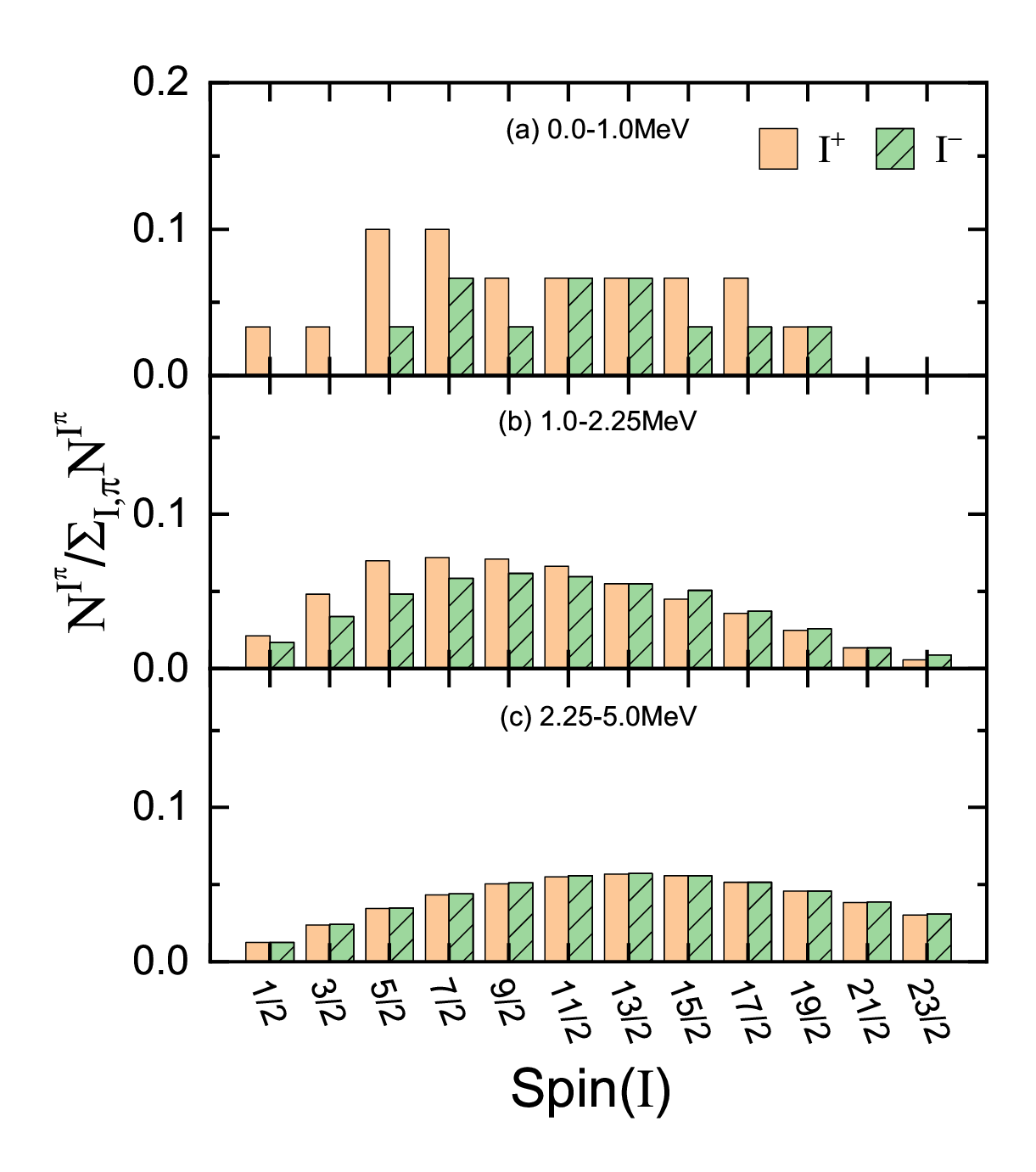} 
\caption{\label{Tb163-spin}(Color online) Histograms showing counts (normalized to unity) separately of odd-parity and even-parity nuclear levels in $^{163}$Tb sorted according to their spin values in the (a) unpaired, (b) pair-breaking, and (c) multi-qp excitation energy regimes.}
\end{figure}

\begin{figure*} \includegraphics[width=6.6in]{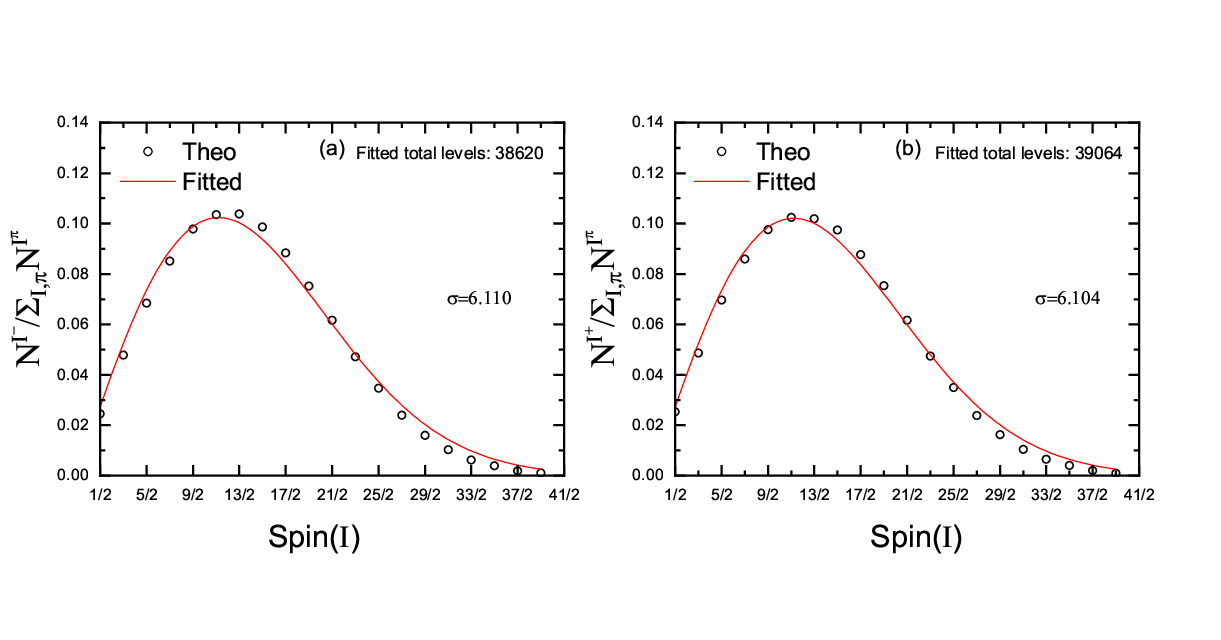}\caption{\label{Dy163-guassfit}(Color online) Least-squares fitting with Ericson's spin-distribution formula (\cite{Ericson1959}) of (a) odd-parity and (b) even-parity levels belonging to the multi-qp regime of $^{163}$Dy (as shown in Fig. \ref{Dy163-spin}(c)).}
\end{figure*}

\begin{figure*}
\includegraphics[width=7in]{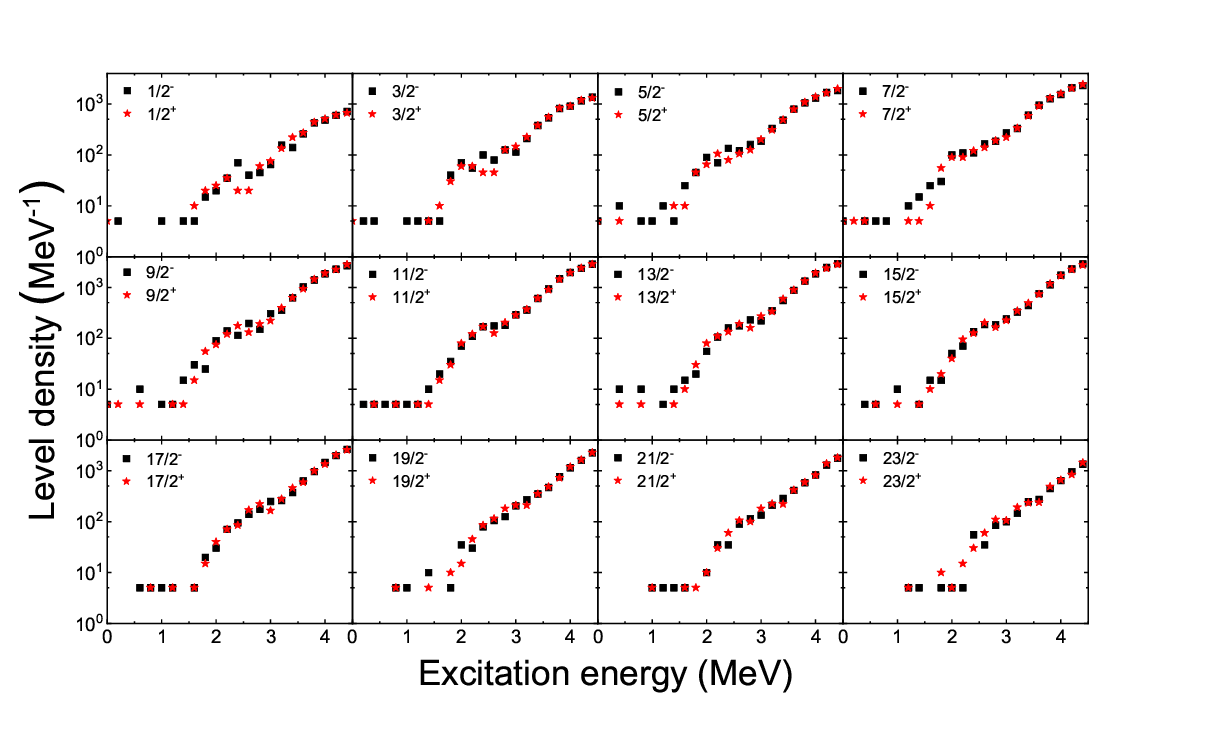} 
\caption{\label{Dy163-sepspin}(Color online) A set of 12 subplots, each representing the density of even-parity (red stars) and odd-parity levels (black squares) in $^{163}$Dy (in the excitation energy bins each being 0.12 MeV wide), those having a definite spin value among the range of 1/2$\hbar$ to 23/2$\hbar$.}
\end{figure*}

After exact angular momentum projection, the calculated states by the PSM are eigenstates of spin. Therefore, we have precise information on the spins for all calculated levels. To see the characteristics of level-spin-distributions at different excitations, we group all the calculated levels in $^{163}$Dy and $^{163}$Tb according to their spins over the three different energy regimes defined in Section \ref{NLDs}. The counts of such-accumulated levels in each of the three energy regimes are normalized to one, decomposed into even and odd parities, and shown in Figs.~\ref{Dy163-spin} and~\ref{Tb163-spin} as functions of spin in the form of histograms. 

The level-spin-distribution in the unpaired regime in even-odd $^{163}$Dy (Fig.~\ref{Dy163-spin}(a)) is clearly different from that in the odd-even $^{163}$Tb (Fig.~\ref{Tb163-spin}(a)). Moreover, remarkable differences can be seen if we compare these two Figures, showing level-spin-distribution in the respective regime of lowest excitations in each of the two odd-$A$ nuclei, with Fig. 5(a) in Ref. \cite{jiaqi2023} showing the same in the lowest excitations (which we termed as the 'collective regime' \cite{jiaqi2023}) in even-even $^{164}$Dy. Interestingly, the feature that has emerged as common among these three nuclei is that, at the lowest excitations, there is by no means a half-half allocation of even- and odd-parity states. Fig.~\ref{Dy163-spin}(a), showing spin-distribution in the unpaired regime in even-odd $^{163}$Dy, features a peak at spin $7/2\hbar$ for levels of both parities and contains more odd-parity states in number than the even-parity ones while Fig.~\ref{Tb163-spin}(a) for odd-even $^{163}$Tb contains more even-parity states and shows a rather irregular spin-distribution for both the parities. Moving up in energy, from the pair-breaking regime in $^{163}$Dy (Fig.~\ref{Dy163-spin}(b)) as well as in $^{163}$Tb (Fig.~\ref{Tb163-spin}(b)), Gaussian-like spin-distributions emerge. Although in the pair-breaking regime in $^{163}$Tb, even-parity states are slightly more in number than the odd-parity ones, the distribution of spins in levels of both parities, in both nuclei, show nearly perfect Gaussian forms. Moreover, shapes of the Gaussian spin-distributions of the odd-parity and the even-parity levels can be seen as almost identical in both odd-$A$ nuclei. Especially, if spin-distributions in the multi-qp regimes of both the nuclei, i.e., Fig.~\ref{Dy163-spin}(c) for even-odd $^{163}$Dy and Fig.~\ref{Tb163-spin}(c) for odd-even $^{163}$Tb are compared, a high level of similarity can be found. This may physically mean that in the multi-qp regime at which the system is dominated by chaotic motion, both odd-$A$ nuclei, one being even-$Z$ and the other being even-$N$, completely forget their characteristic SP structure near the ground state due to which their spin- and parity-dependent behavior becomes almost identical. Therefore, in the following discussion, where we will be analyzing in detail, the Gaussian form of the spin-distribution in the multi-qp regime, it is sufficient to consider only $^{163}$Dy as the representative of odd-$A$ (both even-$N$ or even-$Z$) systems. 

It is straightforward to perform a numerical fitting of the calculated Gaussian spin-distribution shown in Fig.~\ref{Dy163-spin}(c) for $^{163}$Dy with the Ericson's spin-distribution formula \cite{Ericson1959} and determine the desired quantity, the dispersion $\sigma$. We remark that $\sigma$ alone absorbs all the nuclear structure information of the nuclear levels, and calculation of this quantity using a microscopic model is therefore of great interest. We have used the least-squares method available in the MATLAB {\it lsqcurvefit}. In Fig.~\ref{Dy163-guassfit}, in two divisions, namely, (a) and (b), we show the results of fitting of odd- and even-parity levels, respectively, that fall in the energy range 2.75 - 5.0 MeV. These levels belong to the multi-qp regime as we defined in Section \ref{NLDs}. Note that, to draw the long tail in the high-spin part of the Gaussian curve, we extend the projection calculation to $I = 39/2\hbar$. Therefore, as compared to Fig.~\ref{Dy163-spin}(c), Fig.~\ref{Dy163-guassfit} includes more levels: 38,620 levels with odd-parity and 39,064 levels with even-parity. We obtain $\sigma$ = 6.110 and 6.104 for odd- and even-parity, respectively.

Finally, to show the fact that our model can produce NLDs as functions of explicit spins and parities, $\rho(E_x, I,\pi)$ (see Eq.~(\ref{EQ: uncorrelated_NLD}) and related discussion), in Fig.~\ref{Dy163-sepspin}, we present the density of separately, odd-parity and even-parity levels in $^{163}$Dy with fixed spin values among the broad range of spins: 1/2$\hbar$ to 23/2$\hbar$. The result indicates that at a particular excitation energy, $\rho(E_x, I,\pi)$ for levels of different spin values can differ by several factors (even by up to an order of magnitude), and the largest ones in $^{163}$Dy are found for levels of mid-range spins from 11/2$\hbar$ to 17/2$\hbar$ of both parities. For almost all the plots, one can see clear structure characteristics similar to what has been observed in Fig.~\ref{Dy163-major}; i.e., one can see the step-like structures that divide the level densities into three excitation energy regimes. Interestingly, the second step-like structure around 3 MeV is more prominent for NLD curves of lower spins. For higher spin ones, especially for $I=$19/2$\hbar$, 21/2$\hbar$, and 23/2$\hbar$ with odd-parity, the second step structure is smeared out and straight lines show up at very low excitations. This suggests that NLDs of such high-spin states tend to approach the statistical behavior at lower excitation energies than the NLDs of low-spin states.

We note that the subplots in Fig.~\ref{Dy163-sepspin}, each one representing the density of levels of a fixed spin and parity, are made of levels calculated over the entire energy range from 0 to 4.5 MeV. We note that such plots can be readily and consistently created for levels within well-defined energy ranges, whenever needed.

\section{SUMMARY AND FUTURE PROSPECTIVE}\label{Summary}

As a follow-up to our previous work published as Phys. Rev. C {\bf 108}, 034309 (2023) for even-even nuclei, the current article, dealing with odd-$A$ systems, is the second one in our planned series of publications on the shell model study of nuclear level density. In our model, the level density, commonly treated as a statistical quantity, is obtained by solving the quantum-mechanical eigenvalue equation $\hat H\left|\Psi\right> = E \left|\Psi\right>$. Here $\left|\Psi\right>$ is the $many$-$body$ wavefunction, containing all necessary information on the participating particles. Therefore, all energy levels contributing to the density are {\it eigenstates} of spin and parity. 

Pairing and pair breaking have been the central issues in our discussion of the nuclear structure effects on the NLDs. Since the last nucleon in an odd-$A$ system is blocked from participating in the pair formation resulting in a weakened pairing, pronounced differences between NLDs in odd-$A$ and even-even adjoin nuclei have been found. We have discussed in detail that the structure-dominant effects on nuclear levels at low energies, which we noticed to be present strongly in the even-even systems (in our former study \cite{jiaqi2023}), are greatly suppressed in odd-$A$ systems. For example, statistical properties characterized by an equal division of even- and odd-parity levels and Gaussian distribution of level-spins proposed by Ericson begin to appear at much lower excitations in odd-$A$ nuclei than in their even-even adjoins. The realization of a perfect Gaussian spin-distribution in levels with both parities above $\sim$ 2.75 MeV from our calculation eventually permitted us to extract the energy-dependent $\sigma$ values of Ericson's spin-distribution formula for the relevant levels of separate parities. 

An interesting topic that has not been covered within the scope of the present paper, is how to use the nuclear level density as a probe to study the thermodynamic properties of nuclei. In Ref. \cite{Nyhus2012}, Nyhus {\it et al.}, using their Oslo data, showed that thermodynamic properties of nuclei can be deduced using both micro-canonical and canonical ensemble theories. We have the full set of many-body wavefunction $\left|\Psi\right>$ at our disposal that can, in principle, be applied to calculate any thermodynamic quantities. One may talk about a phase transition from the pair-correlated state at low energies to maybe, new excitation modes at high energies, as conjectured by Voinov {\it et al.} \cite{Voinov2009}. The nature of such excitation modes remains open for explanation, but cannot be understood using the non-interacting Fermi gas model, as suggested in Ref. \cite{Moretto2015}. In fact, our shell model, which, as shown here, can essentially reproduce the level density data known from experiments, treats the high-energy states as {\it strongly-interacting} multi-qp states that differ fundamentally from the non-interacting Fermi-gas states.
 
Little has been discussed about the nuclear level density in odd-odd systems. Odd-odd nuclei, in which both types of fermionic particles are blocked from pairing in the ground state and in which, the lowest configurations are the 2-qp states formed from the coupling of an unpaired neutron with an unpaired proton, may exhibit yet unknown properties of levels. Shell model calculations to obtain energy states in odd-odd nuclei are in progress and results will be reported in a future publication. 

\begin{acknowledgments}
Valuable discussions with M. Wiedeking and A. V. Voinov are acknowledged. We also thank the discussions with all the participants in the {\it First International Workshop on Neutron Capture Reactions} held in November 2023 at Huizhou, China. This work is supported by the National Natural Science Foundation of China (NSFC) (Grant Nos. 12235003, 12275225), by the Special Foundation for Theoretical Physics, NSFC (Grant No. 12347125), and by the China Postdoctoral Science Foundation (Grant Nos. 2023TQ0217,  2024M751951).
\end{acknowledgments}


\begin{thebibliography}{99}

\bibitem{Larsen2019} A. C. Larsen, A. Spyrou, S. N. Liddick, and M. Guttormsen, Prog. Part. Nucl. Phys. {\bf 107}, 69 (2019).

\bibitem{Hauser-Feshbach} W. Hauser and H. Feshbach, Phys. Rev. {\bf 87}, 366 (1952).

\bibitem{Oslo} A. Schiller {\it et al.}, Nucl. Instrum. Methods Phys. Res. A {\bf 447}, 494 (2000).

\bibitem{Beta-Oslo} A. Spyrou {\it et al.}, Phys. Rev. Lett. {\bf 113}, 232502 (2014).

\bibitem{Inverse-Oslo} V. W. Ingeberg {\it et al.}, Eur. Phys. J. A {\bf 56}, 68 (2020).

\bibitem{Goriely_EPJA}S. Goriely {\it et al.}, Eur. Phys. Jour. A {\bf55}, 172 (2019).

\bibitem{Shape method} M. Wiedeking {\it et al.}, Phys. Rev. C {\bf 104}, 014311 (2021).

\bibitem{Shape_method_1}D. M{\"u}cher {\it et al.}, Phys. Rev. C {\bf107}, L011602 (2023).

\bibitem{Ericson1959}T. Ericson, Nucl. Phys. {\bf 11}, 481 (1959).

\bibitem{Bethe1937} H. A. Bethe, Rev. Mod. Phys. {\bf 9}, 69 (1937).

\bibitem{Gilbert-Cameron}A. Gilbert and A. G. W. Cameron, Can. J. Phys. {\bf43}, 1446 (1965).

\bibitem{combinatorial_hfb_0}S. Hilaire and S. Goriely,  Nucl. Phys. A {\bf 779}, 63 (2006).
 
\bibitem{combinatorial_hfb_1}S. Goriely, S. Hilaire, and A. J. Koning, Phys. Rev. C {\bf 78}, 064307 (2008).

\bibitem{Ring-Schuck} P. Ring and P. Schuck, {\it The nuclear many-body problem}, (Springer Verlag, Berlin, 2004).

\bibitem{monte-carlo-shell-model} G. H. Lang, C. W. Johnson, S. E. Koonin, and W. E. Ormand, Phys. Rev. C {\bf48}, 1518 (1993).

\bibitem{smmc_2} Y. Alhassid, D. J. Dean, S. E. Koonin, G. Lang, and W. E. Ormand, Phys. Rev. Lett. {\bf72}, 613 (1994).

\bibitem{smmc_3} S. E. Koonin, D. J. Dean, and K. Langanke, Phys. Rep. {\bf278}, 1 (1997).

\bibitem{Alhassid2007} Y. Alhassid, S. Liu, and H. Nakada, Phys. Rev. Lett. {\bf 99}, 162504 (2007).

\bibitem{jiaqi2023}J.-Q. Wang, S. Dutta, L.-J. Wang, and Y. Sun, Phys. Rev. C {\bf108}, 034309 (2023).

\bibitem{PSM_Hara_Sun}K. Hara and Y. Sun, Int. J. Mod. Phys. E {\bf4}, 637 (1995).

\bibitem{PSM_Sun}Y. Sun, Phys. Scr. {\bf91}, 043005 (2016).

\bibitem{data-Dy163} H. H. Schmidt {\it et al.}, Nucl. Phys. A {\bf 504}, 1 (1989). 

\bibitem{smmc_odd_mass} A. Mukherjee and Y. Alhassid, Phys. Rev. Lett. {\bf109}, 032503 (2012).

\bibitem{smmc_rare_earth_odd-even} C. {\"O}zen, Y. Alhassid, and H. Nakada, Phys. Rev. C {\bf91}, 034329 (2015)

\bibitem{Hara1992}K. Hara and Y. Sun, Nucl. Phys. A {\bf 537}, 77 (1992).

\bibitem{Sun1997}Y. Sun and K. Hara, Comp. Phys. Commun. {\bf 104}, 245 (1997).

\bibitem{Sun1994a} Y. Sun, S. X. Wen, and D. H. Feng, Phys. Rev. Lett. {\bf 72}, 3483 (1994).

\bibitem{Sun1994b} Y. Sun, D. H. Feng, and S. X. Wen, Phys. Rev. C {\bf 50}, 2351 (1994).

\bibitem{Ta177-Exp} D. E. Archer {\it et al.}, Phys. Rev. C {\bf 52}, 1326 (1995).

\bibitem{Wang_2018_GT_rates} L.-J. Wang, Y. Sun, and S. K. Ghorui, Phys. Rev. C {\bf97}, 044302 (2018).

\bibitem{urca_cooling} L.-J. Wang, L. Tan, Z.-P. Li, B. Gao, and Y. Sun, Phys. Rev. Lett. {\bf127}, 172702 (2021).

\bibitem{EC_rates} L. Tan, Y.-X. Liu, L.-J. Wang, Z. Li, and Y. Sun,  Phys. Lett. B {\bf805}, 135432 (2020).

\bibitem{Nil-1985}T. Bengtsson and I. Ragnarsson, Nucl. Phys. A {\bf436}, 14 (1985).

\bibitem{Guttormsen2000}M. Guttormsen  {\it et al.}, Phys. Rev. C {\bf 61}, 067302 (2000).

\bibitem{NNDC} NNDC database at https://www.nndc.bnl.gov/nudat3/

\bibitem{oslo_Midtbo}J. E. Midtb{\o} {\it et al.}, Comp. Phys. Comm. {\bf262}, 107795 (2021). 

\bibitem{Oslo-Dy163} T. Renstr{\o}m {\it et al.}, 
Phys. Rev. C {\bf 98}, 054310 (2018).

\bibitem{Zelevinsky2019} V. Zelevinsky and M. Horoi, Prog. Part. Nucl. Phys. {\bf 105}, 180
(2019).

\bibitem{Guttormsen2000b} M. Guttormsen {\it et al.}, Phys. Rev. C {\bf 62}, 024306 (2000). 

\bibitem{Ericson1960} T. Ericson, Adv. Phys. {\bf9}, 425 (1960).

\bibitem{Huizenga1972}J. Huizenga and L. G. Moretto, Ann. Rev. Nucl. Sci. {\bf22}, 427 (1972).

\bibitem{Mocelj2005} D. Mocelj  {\it et al.}, Nucl. Phys. A {\bf 758}, 154c (2005).

\bibitem{Alhassid2000} Y. Alhassid, G. F. Bertsch, S. Liu, and H. Nakada, Phys. Rev. Lett. {\bf 84}, 4313 (2000).

\bibitem{Herman1987} M. Herman and G. Reffo, Phys. Rev. C {\bf 36}, 1546 (1987).

\bibitem{Herman1988} M. Herman, G. Reffo, and R. A. Rego, Phys. Rev. C {\bf 37}, 797 (1988).

\bibitem{Cerf1991} N. Cerf, Phys. Lett. B {\bf 268}, 317 (1991).

\bibitem{Cerf1993} N. Cerf, Nucl. Phys. A {\bf 554}, 85 (1993).

\bibitem{sensitivity_1}S. N. Liddick  {\it et al.}, 
Phys. Rev. Lett. {\bf116}, 242502 (2016).

\bibitem{Fission_HFMC_1} B. Becker, P. Talou, T. Kawano, Y. Danon, and I. Stetcu, Phys. Rev. C {\bf 87}, 014617 (2013).

\bibitem{Fission_HFMC_2} P. Talou, T. Kawano, and I. Stetcu, Phys. Procedia {\bf47}, 39 (2013). 

\bibitem{fission_monte-carlo} I. Stetcu, P. Talou, T. Kawano, and M. Jandel, Phys. Rev. C {\bf90}, 024617 (2014).

\bibitem{fission_monte-carlo_2} I. Stetcu, P. Talou, T. Kawano, and M. Jandel, Phys. Rev. C {\bf88}, 044603 (2013).

\bibitem{Voinov2007} A. V. Voinov {\it et al.}, Phys. Rev. C {\bf 76}, 044602 (2007). 

\bibitem{Grimes2013} S. M. Grimes, Phys. Rev. C {\bf 88}, 024613 (2013).

\bibitem{Grimes2016} S. M. Grimes, A. V. Voinov, and T. N. Massey, Phys. Rev. C {\bf 94}, 014308 (2016).

\bibitem{Nyhus2012} H. T. Nyhus {\it et al.}, Phys. Rev. C {\bf 85}, 014323 (2012).

\bibitem{Voinov2009} A. V. Voinov {\it et al.}, Phys. Rev. C {\bf 79}, 031301(R) (2009).


\bibitem{Moretto2015} L. G. Moretto, A. C. Larsen, F. Giacoppo, M. Guttormsen, and S. Siem, J. Phys. Conf. Ser. {\bf 580}, 012048 (2015).

\end{thebibliography}
\end{document}